\documentclass[11pt]{article}
\pdfoutput=1 
\usepackage{jheppub}
\usepackage{graphicx,verbatim}
\usepackage{psfrag, times}
\usepackage[T1]{fontenc}
\usepackage{epsf}

\def\be{\begin{equation}}
\def\ee{\end{equation}}
\def\bc{\begin{center}}
\def\ec{\end{center}}

\def\bR{{\mathbb{R}}}

\def\bM{{\mathbb{M}}}
\def\bbP{{\mathbb{P}}}

\def\bbQ{{\mathbb{Q}}}
\def\bbC{{\mathbb{C}}}
\def\bbA{{\mathbb{A}}}
\def\bbT{{\mathbb{T}}}

\def\cA{{\mathcal{A}}}
\def\cB{{\mathcal{B}}}
\def\cC{{\mathcal{C}}}
\def\cD{{\mathcal{D}}}
\def\cH{{\mathcal{H}}}

\def\cF{{\mathcal{F}}}

\def\cN{{\mathcal{N}}}

\def\cS{{\mathcal{S}}}
\def\cT{{\mathcal{T}}}
\def\cI{{\mathcal{I}}}
\def\cJ{{\mathcal{J}}}

\def\cK{{\mathcal{K}}}

\def\cO{{\mathcal{O}}}
\def\cP{{\mathcal{P}}}
\def\cV{{\mathcal{V}}}
\def\cW{{\mathcal{W}}}
\def\cY{{\mathcal{Y}}}

\def\tX{\tilde{X}}
\def\tW{\tilde{W}}
\def\pa{\partial}

\def\ga{\gamma}
\def\nn{\nonumber}
\def\lam{\lambda}

\def\Del{{\Delta}}

\def\r2{{\sqrt{2}}}

\def\bea{\begin{eqnarray}}
\def\eea{\end{eqnarray}}
\def\ru{{\rm{u}}}
\def\rn{{\rm{n}}}
\def\cT{{\mathcal{T}}}

\def\tA{\tilde{A}}
\def\tB{\tilde{B}}

\def\tX{\tilde{X}}
\def\tK{{\tilde{K}}}
\def\tDel{\tilde{\Del}}
\def\rD{{\rm D}}
\def\rH{{\rm H}}

\def\rV{{\rm V}}

\def\rC{{\rm C}}

\newcommand{\no}{\nonumber}

\begin{document}

\begin{flushright}
KUNS-2664
\end{flushright}

\title{\Large Anatomy of Geodesic Witten Diagrams}
\author[]{Heng-Yu Chen${}^{1}$, En-Jui Kuo${}^{1}$ and Hideki Kyono${}^{2}$}
\affiliation{$^1$\rm Department of Physics and Center for Theoretical Sciences, \\
National Taiwan University, Taipei 10617, Taiwan}
\affiliation{$^2$ \rm Department of Physics, Kyoto University,
Kitashirakawa Oiwake-cho, Kyoto 606-8502, Japan}

\emailAdd{heng.yu.chen@phys.ntu.edu.tw}
\emailAdd{r04222087@ntu.edu.tw}
\emailAdd{h\_kyono@gauge.scphys.kyoto-u.ac.jp}
\abstract{We revisit the so-called ``Geodesic Witten Diagrams'' (GWDs)  \cite{ScalarGWD}, proposed to be the holographic dual configuration of scalar conformal partial waves, from the perspectives of CFT operator product expansions. To this end, 
we explicitly consider three point GWDs which are natural building blocks of all possible four point GWDs, discuss their gluing procedure through integration over spectral parameter,
and this leads us to a direct identification with the integral representation of CFT conformal partial waves. As a main application of this general construction, we consider the holographic dual of the conformal partial waves for external primary operators with spins. Moreover, we consider the closely related ``split representation'' for the bulk to bulk spinning propagator, to demonstrate how ordinary scalar Witten diagram with arbitrary spin exchange, can be systematically decomposed into scalar GWDs. We also discuss how to generalize to spinning cases.
}

\maketitle

\section{Introduction}\label{Section:Introduction}
\paragraph{}
One of the most powerful applications of AdS/CFT correspondence is that we can realize the important and sometimes complicated CFT observables such as correlation functions, 
through computationally simple geometric configurations inside the dual Anti-de Sitter space (See \cite{correlator1}, \cite{correlator2} for selected references, and \cite{AdSCFT-review1} for a good review in this area.). 
Such an application often relies heavily on the underlying conformal symmetries or equivalently the isometries of Anti-de Sitter space.
Conformal blocks, which allow us to disentangle what are universally constrained by conformal symmetries in four point CFT correlation functions,
from theory-dependent data, such as spectrum of scale dimensions $\{\Del_i\}$ and OPE coefficients $\{\lambda_{ijk}\}$, 
offer a ideal venue for such a geometric realization in the dual AdS space.
\paragraph{}
Curiously, despite almost twenty years since the inception of AdS/CFT correspondence, 
the holographic dual configuration  of conformal block, termed ``geodesic Witten diagram'' (GWD), have only been constructed recently in a striking paper \cite{ScalarGWD}.
In a complete analogy with the CFT decomposition, 
the ordinary scalar four point Witten diagrams which holographically computes the full four point CFT correlation functions,
can be shown to decompose into a summation over the GWDs.
Moreover, each of these scalar GWDs involved in the sum, can be identified directly with the conformal block for single and double trace primary operator exchange.
\paragraph{}
However from the perspective of CFT
operator product expansions, it is sometimes more illuminating to think instead about the individual conformal block $G_{\Del, J}(u, v)$ as being built from fusing a pair of three point functions, 
each involving two of the external primary operators and the internal exchange operator $\cO_{\Del, J}$ itself (For good recent CFT reviews, see \cite{CFT-review1, CFT-review2} ). 
Indeed this fusion procedure of three point functions was made explicit in \cite{DO-2011} (and later extended in \cite{DSD-Shadow}), 
through defining the so-called ``shadow operators'',
which yields the integral representation of conformal block. This will be reviewed in the next section. 
It is therefore natural to ask if this further decomposition procedure of individual conformal blocks themselves can also be seen in AdS space, 
perhaps directly cutting up a four point GWD in the middle into two   
three point ones? It turns out that this intuitive picture is qualitatively correct, and the detailed justification comes from the non-trivial identity between the bulk to bulk and bulk to boundary propagators we shall derive. 
We shall name the resulting building element: three point geodesic Witten diagram, see Figure \ref{Fig:3pt-geostraight}. 
The main difference from the ordinary three point Witten diagram is now that the bulk interaction point is restricted to move along the geodesic connecting two our of three boundary points.
As we will see in Section \ref{Section:Review}, this procedure of cutting and rejoining also allows us to directly identify four point scalar GWD with the integral representation of scalar conformal block by construction, hence provides an alternative proof for the results in \cite{ScalarGWD}.
\paragraph{}
As an main application of this understanding, the three point GWDs become particularly useful when constructing the holographic dual of spinning conformal blocks \cite{SpinningBlock0, SpinningBlock1}, as they allow us to directly apply the earlier general parameterization of three point vertex for three symmetric traceless tensor fields constructed in \cite{3pt-Coupling, 3pt-Coupling0}  (up to certain modification to account for the restriction along the geodesic) to study the precise nature of the interaction. The resultant calculations can then be expressed in terms of appropriate CFT tensor structures, we will provide explicit examples and illustrate how general spinning geodesic Witten diagrams can be constructed in Section \ref{Section:AdS}. We will review the relevant CFT details in Section \ref{Sec:SpinningCFT}.
\paragraph{}
Finally, the analysis we have done is closely related to the so-called ``Split representation'' of bulk to bulk propagator \cite{SpinningAdS, SpinningAdS2}.
in fact we will demonstrate its power by combining with the knowledge of three point GWDs in Section \ref{Section:Decomposition}. 
Explicitly we can rewrite the split representation of four point scalar Witten diagrams with arbitrary spin-$J$ exchange 
into a summation over products of three point GWDs. 
By explicitly identifying the physical residues when performing the integration over so-called ``spectral parameter'' $\nu$,
we can show that the summation contains one four point scalar GWD for single trace operator with spin-$J$,   
plus infinite towers of four point scalar GWDs for double trace primary operators with spins $0, 1, \dots, J$. This is consistent with and completes analysis in \cite{ScalarGWD} for $J=0, 1$ cases.
We also discuss how similar decompositions can be done for spinning Witten diagrams into spinning GWDs.
\paragraph{}
We relegate some useful background materials and computational details into several appendices.
\paragraph{}
While this work is being finalized, two nice preprints \cite{Castro1}, \cite{Dyer1} appeared\footnote{
{Another nice paper \cite{Sleight2017} has appeared simultaneously when we submitted version 1 of this work to arXiv, and their work also has some partial overlaps.}}, 
which have partial overlaps with our results.
However we hope our independent work, which has somewhat different computational approaches and topical emphases, can complement their works. 
An earlier work \cite{Nishida1}, which considered holographic dual of conformal block with single external operator with spin, also contained a special case of our results\footnote{Please also see \cite{OPEblock} for the interesting connections between so-called ``OPE blocks'' and geodesic Witten diagrams.}.

\section{Scalar Four Point Geodesic Witten Diagrams Revisited}\label{Section:Review}
\paragraph{}
Let us begin by reviewing the essential details about the geodesic Witten diagram in $d+1$ dimensional Anti-de Sitter space AdS$_{d+1}$ \cite{ScalarGWD}.
This was proposed to be the holographic dual configuration of the $d$-dimensional scalar conformal partial wave associated with the exchange of a primary operator $\cO_{\Del, J}$ of scaling dimension $\Del$ and spin $J$ and its conformal descendants between two pairs of external local scalar primary operators $\cO_{\Del_{1}}(P_{1})$,  $\cO_{\Del_{2}}(P_{2})$  
and $\cO_{\Del_{3}}(P_{3})$,   $\cO_{\Del_{4}}(P_{4})$: 
\be\label{Def:ScalarCPW}
W_{\cO_{\Del, J}}(P_1, P_2, P_3, P_4) = \left(\frac{P_{24}}{P_{14}}\right)^{\frac{\Del_{12}}{2}} \left(\frac{P_{14}}{P_{13}}\right)^{\frac{\Del_{34}}{2}} \frac{G_{\cO_{\Del, J}}(u, v)}{(P_{12})^{\frac{\Del_1+\Del_2}{2}} (P_{34})^{\frac{\Del_3+\Del_4}{2}} },
\ee
where $\Del_{ij} =\Del_i-\Del_j$.
In this note we will mostly use so-called ``embedding formalism'' reviewed in Appendix \ref{Appendix:Embed} and follow the conventions in \cite{CFT-review1}. 
Here $P_i$ labels the position of operator $\cO_{\Del_i}(P_i)$ in $d+2$ dimensional embedding space, and their separations are:
\be\label{Def:Pij}
P_{ij} = -2 P_i \cdot P_j, \quad i,j=1,2,3, 4.
\ee
We can express the ``scalar conformal block'' $G_{\cO_{\Del, J}}(u, v)$ for $\cO_{\Del, J}$ as a function of the two independent conformally invariant cross-ratios:
\be\label{Def:cross ratio}
u = \frac{P_{12} P_{34}}{P_{13}P_{24}}, \quad v=\frac{P_{14}P_{23}}{P_{13}P_{24}}.
\ee
The closed form expressions of $G_{\cO_{\Del, J}}(u, v)$ for even $d$-dimensions have been solved explicitly in terms of hypergeometric functions using quadratic Casimir operators \cite{DO-2003, DO-2011};
more recently the precise connections of $G_{\cO_{\Del, J}}(u, v)$ with the eigenfunctions of quantum integrable systems have also been established for arbitrary $d$-dimensions in \cite{Integrability1, Integrability2}.
\begin{figure}[h]\centering
\includegraphics[width=7.5cm]{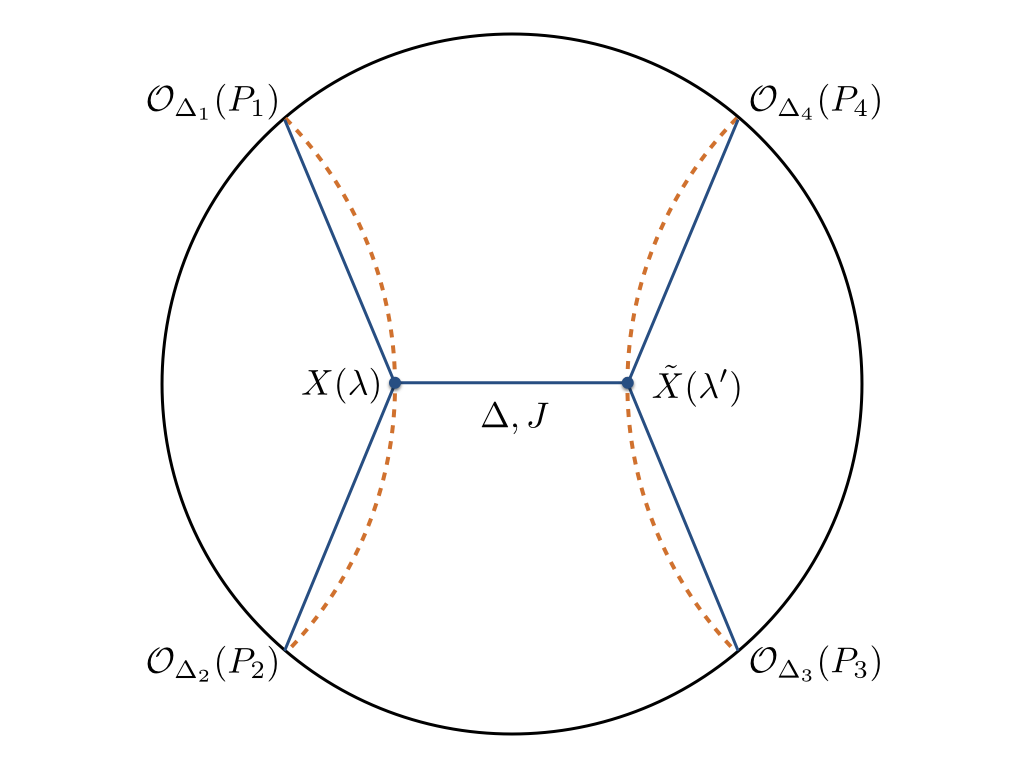}
\caption{{Four point scalar geodesic Witten diagram where the orange curves describe the geodesics and the blue lines are bulk to boundary propagators, such that the interaction vertices move along geodesics $\gamma_{12}$ and $\gamma_{34}$.}  \label{Fig:4pt-ScalarGWD}
  }
\end{figure}
\paragraph{} 
Now imagine these external scalar primary operators $\{\cO_{\Del_i}\}$ are inserted at the boundary of AdS$_{d+1}$ at points $\{P_i\}$, 
and use $\ga_{12}$ and $\ga_{34}$ to denote the geodesics connecting the points $P_{1,2}$ and $P_{3,4}$ respectively, 
the four point scalar geodesic Witten diagram is defined through the double integral (See Figure \ref{Fig:4pt-ScalarGWD}): 
\be\label{Def:ScalarGWD1}
\cW_{\Del, J}(P_i)=
 \int_{\ga_{12}} \int_{\ga_{34}}\prod_{c=1}^2\Pi_{\Del_c}(X(\lam), P_c)\, \hat{\Pi}_{\Del, J}\left(X(\lam), \tX(\lam'); \frac{dX(\lam)}{d\lam}, \frac{d\tX(\lam')}{d\lam'} \right)\,
\prod_{c'=3}^4 \Pi_{\Del_{c'}}(\tX(\lam'), P_{c'})\,.
\ee
Here $-\infty < \lambda, \lambda' < +\infty$ are the line parameters of $\ga_{12}$ and $\ga_{34}$ which we integrate along with, 
in terms of bulk AdS$_{d+1}$ coordinates $X^A(\lam)$ and $\tX^A(\lam')$, the two geodesics are given by following curves:
\be\label{Def:X-geodesics}
\ga_{12}~:~X^A(\lam) = \frac{P_1^A e^\lam+P_2^A e^{-\lam}}{{(P_{12})}^{\frac{1}{2}}}, \quad\quad  \ga_{34}~:~\tX^A(\lam') = \frac{P_3^A e^{\lam'}+P_4^A e^{-\lam'}}{{(P_{34})}^{\frac{1}{2}}}.
\ee
The integrand in \eqref{Def:ScalarGWD1} consists of the pull-back of  bulk to boundary scalar propagators\footnote{Here the overall normalization constant $\cC_{\Del} =\frac{\Gamma(\Del)}{2\pi^{\frac{d}{2}}\Gamma(\Del+1-\frac{d}{2})}$ is defined as a special case of \eqref{Def:CDel-l}.}:
\be\label{Def:scalarPiBb1234}
\Pi_{\Del_{1,2}}(X(\lam), P_{1,2}) = \frac{\cC_{\Del_{1,2}}}{(-2P_{1,2}\cdot X(\lam))^{\Del_{1,2}}}, 
\quad
\Pi_{\Del_{3,4}}(\tX(\lam'), P_{3,4}) = \frac{\cC_{\Del_{3,4}}}{(-2P_{3,4}\cdot\tX(\lam'))^{\Del_{3,4}}}, 
\ee
and the pull-back of bulk to bulk propagator of spin-$J$ tensor field between $X^A(\lam)$ and $\tX^A(\lam')$ on the two geodesics:
\be\label{Def:SpinJ-prop}
\hat{\Pi}_{\Del, J}\left(X(\lam), \tX(\lam'); \frac{dX(\lam)}{d\lam}, \frac{d\tX(\lam')}{d\lam'} \right) = \left(\frac{d X(\lam)}{d\lam}\right)_{A_1, \dots A_J}\left(\frac{d \tX(\lam')}{d\lam'}\right)_{\tA_1,\dots, \tA_J}
\Pi_{\Del, J}^{\{A_1\dots A_J\},\{ \tilde{A}_1\dots \tilde{A}_J\} } (X, \tX ). 
\ee 
Here $\Pi_{\Del, J}^{\{A_1\dots A_J\},\{ \tilde{A}_1\dots \tilde{A}_J\} } (X, \tX )$ is a (doubly) symmetric, traceless and transverse (STT) tensor whose form will be specified momentarily, 
such that each set of indices satisfy $X_{A_1} \Pi^{\{A_1 A_2 \dots A_J\}} = 0$ and $\eta_{A_1 A_2} \Pi^{\{A_1, A_2\dots A_J\}} = 0$. 
In \eqref{Def:SpinJ-prop}, we have introduce the short-hand notation:
\be\label{SH-notation}
Y^{A_1\dots A_J} \equiv Y^{A_1}\dots Y^{A_J}
\ee
to denote symmetric tensor built from the products of identical vector or vectorial operator $Y^A$.
The proposal of geodesic Witten diagram \cite{ScalarGWD} 
is such that instead of integrating the bulk interacting vertices $(X, \tX)$ over the entire AdS$_{d+1}$ as in computing the holographic correlation functions,
they are pulled back to move only along the geodesic trajectories \eqref{Def:X-geodesics} and the integration in \eqref{Def:ScalarGWD1} is taken along the line parameters $\lambda$ and $\lambda'$.
By showing \eqref{Def:ScalarGWD1} satisfies the eigenvalue equation of quadratic conformal Casimir operator, the authors of \cite{ScalarGWD} explicitly established:
\be\label{Result1}
\cW_{\Del, J}(P_i) \equiv W_{\cO_{\Del, J}}(P_i)
\ee
up to an unimportant overall normalization constant, in our subsequent computations, we will do the same unless otherwise stated.
Moreover, we will provide an alternative proof for \eqref{Result1} by considering three point geodesic Witten diagrams momentarily.

\begin{figure}[h]\centering
\includegraphics[width=7.5cm]{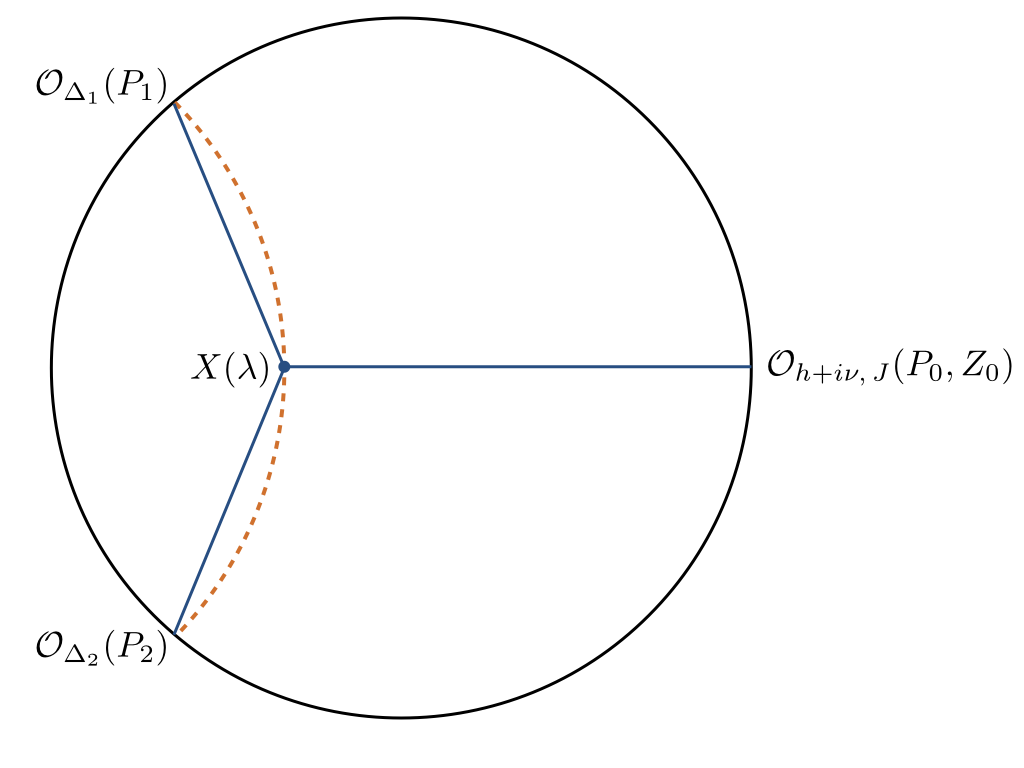}
\caption{Three point scalar geodesic Witten diagram  \label{Fig:3pt-geostraight}
  }
\end{figure}

\paragraph{}
The doubly STT tensor  
$\Pi_{\Del, J}^{\{A_1\dots A_J\},\{ \tilde{A}_1\dots \tilde{A}_J\}}(X,\tX)$ in \eqref{Def:SpinJ-prop} can be obtained from the following index-free generating polynomial
\cite{SpinningAdS}:
\be\label{Def:spinJPibb}
\Pi_{\Del, J}(X, \tX; W, \tW) = \sum_{k=0}^{J}(W\cdot\tW)^{J-k}\left((W\cdot\tX)(\tW\cdot X)\right)^k g_k(\ru), \quad \ru = -1-X\cdot \tX
\ee
where $W^A$ (and $\tW^A$) is the auxiliary polarization vector satisfying $W\cdot X = W\cdot W = 0$ 
and the function $g_k(\ru)$ can be explicitly obtained from the equation of motion for a massive spin $J$ particle in terms of hypergeometric functions.
Next we act on \eqref{Def:spinJPibb} with the product of projection operators $K_A$ and $\tK_{\tA}$:
\bea\label{Def:spinJPibb-components}
&&\Pi_{\Del, J}^{\{A_1\dots A_J\},\{ \tilde{A}_1\dots \tilde{A}_J\} } (X, \tX )= \frac{1}{[J!(\frac{d-1}{2})_J]^2} K^{A_1\dots A_J}\tK^{\tA_1\dots \tA_J} \Pi_{\Del, J}(X, \tX; W, \tW )\nn\\
&=&\sum_{k=0}^{J} G^{\{ A_1}_{B_1}\dots G^{ A_J\}}_{B_J} \tilde{G}^{\{ \tA_1}_{\tB_1}\dots \tilde{G}^{ \tA_J\}}_{\tB_J}\eta^{B_1\tB_1} \dots \eta^{B_{J-k}\tB_{J-k}} \tX^{B_{J-k+1}}\dots \tX^{B_J}
X^{\tB_{J-k+1}}\dots X^{\tB_J} g_k(\ru)\,,
\nn\\
\eea
where the Pochhammer symbol is defined to be $(x)_J = \frac{\Gamma(x+J)}{\Gamma(x)}$.
The explicit form of $K_A$ is given in \eqref{Def:KAop}, it satisfies $K_A K_B =K_B K_A$ (symmetric), $K^A K_A  =0$ (traceless) and $X^A K_A = 0$ (transverse),
it allows us to implement the contraction between various STT tensors before we restrict to geodesics.
Here we have also introduced the induced AdS metric $G_{AB}$  and the projection operator $G_A{}^B$ in the embedding space:
\be\label{Def:GAB}
G_{AB} = \eta_{AB} + X_A X_B,  \quad   G_A{}^{B} =\delta_A^B+X_A X^B,\quad G_A{}^C G_C{}^B =G_A{}^B, \quad G_{A}{}^B X_B = X^A G_A{}^B=0.
\ee
When contracting product of $G_A{}^B$ with an arbitrary tensor in embedding space, such a tensor is then projected into the one satisfying the transverse condition, 
hence in the interior of hyperboloid corresponding to AdS$_{d+1}$. Identical quantities can be defined for the other bulk vertex point with $X\to \tX$ and $i=1,2 \to i=3,4$.
We can see that under the action of $K_A$ operators,  \eqref{Def:spinJPibb-components} automatically satisfy the symmetric, traceless and transverse conditions.
\begin{figure}[h]\centering
\includegraphics[width=7.5cm]{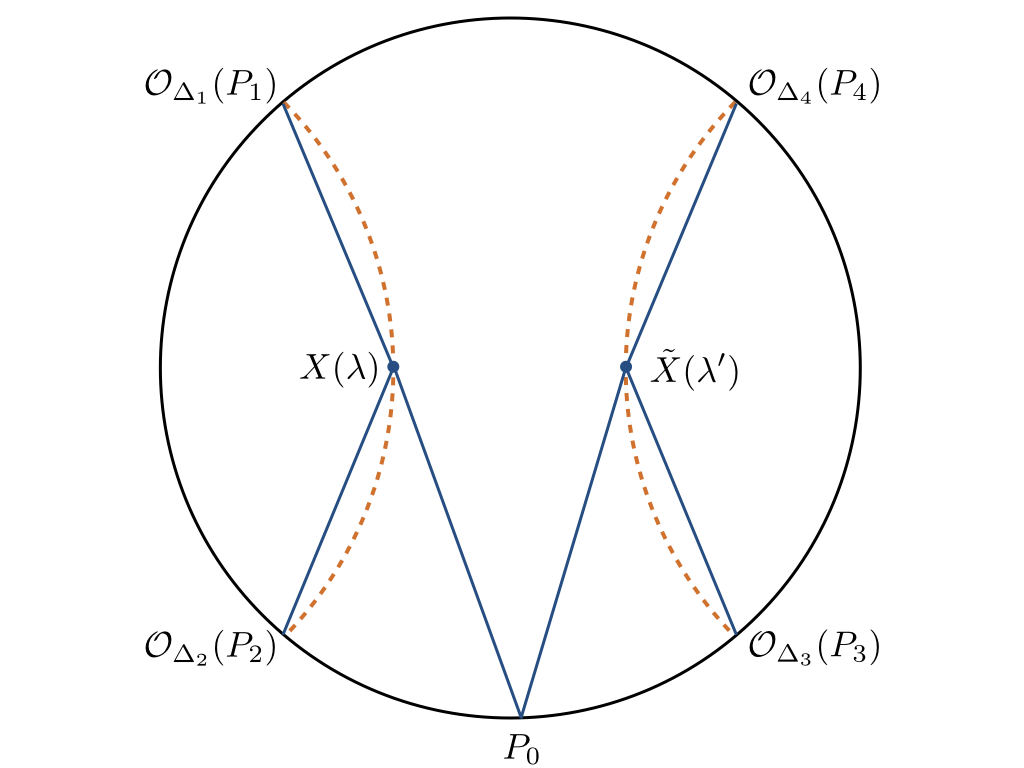}
\caption{Cutting the four point scalar geodesic Witten diagram into the three point ones. \label{Fig:4pt-geosplit}
  }
\end{figure}
\paragraph{} 
It was shown in \cite{SpinningAdS} that the bulk to bulk propagator can be related to the harmonic function $\Omega_{\Del, J}(X, \tX; W, \tilde{W})$ in AdS space as:
\be\label{Def:AdSHarm0}
\Omega_{\nu, J}(X, \tX; W, \tilde{W}) = \frac{i\nu}{2\pi }\left(\Pi_{h+i\nu, J}(X, \tX; W, \tW) - \Pi_{h-i\nu, J}(X, \tX; W, \tW)  \right)
\ee
where $h =\frac{d}{2}$. We can invert this relation by considering following integral identity:
\bea\label{Def:Id-Pi}
 \Pi_{\Del, J}(X, \tX; W, \tW) &=&
\int^{+\infty}_{-\infty}d\nu \frac{1}{\nu^2+(\Del-h)^2}  \Omega_{\nu, J}(X, \tX; W, \tilde{W})\nn\\  
&=&  \int^{+\infty}_{-\infty}\frac{d\nu}{2\pi i} \frac{\nu}{\nu^2+(\Del-h)^2  } \left(\Pi_{h-i\nu, J}(X, \tX; W, \tW) - \Pi_{h+i\nu, J}(X, \tX; W, \tW)  \right),\nn\\
\eea
where in the last line as $\Pi_{h\mp i\nu, J}(X,\tX; W, \tW)$ only converges for ${\rm Im}(\nu) \to \pm \infty$, 
we have closed the integration contour in the upper (lower) half complex $\nu$-plane for first (second) term of the integrand.
Moreover it was shown in \cite{SpinningAdS} that it also admits following representations in terms of bulk to boundary propagators:
\be\label{Def:AdSHarm}
\Omega_{\nu, J}(X, \tX; W, \tilde{W}) = \frac{\nu^2}{\pi J! (h-1)_J}\int_{\partial} dP_0\Pi_{h+iv, J}(X, P_0; W, D_{Z_0}) \Pi_{h-iv, J}(\tX, P_0; \tilde{W}, Z_0).
\ee
Here the spin-$J$ bulk to boundary propagator is:
\be\label{Def:Spin l-bulk-boundary}
\Pi_{\Del, J}(X, P; W, Z) = \cC_{\Del, J}\frac{(2(X\cdot Z)(P\cdot W)-2(X\cdot P)(Z \cdot W))^{J}}{(-2P \cdot X)^{\tau}} 
= \cC_{\Del, J}\frac{(2 X\cdot \rC \cdot W)^{J}}{(-2P\cdot X)^{\tau}},
\ee
where $\tau=\Del+J$ and for later purpose we have also defined the boundary anti-symmetric tensor
\be\label{Def:CiAB}
\rC^{AB} = Z^A P^B - P^A Z^B, \quad Z \cdot P= Z \cdot Z =0 
\ee 
with $Z^A$ being the auxiliary polarization vector associated with boundary point $P^A$.
Notice that $\rC_{AB}$ hence bulk to boundary propagator  \eqref{Def:Spin l-bulk-boundary}  is manifestly invariant under the shift $Z^A \to Z^A+\alpha P^A $.
The overall normalization constant is fixed to be:
\be\label{Def:CDel-l}
\cC_{\Del, J} = \frac{(J+\Del-1)\Gamma(\Del)}{2\pi^{h}(\Del-1)\Gamma(\Del+1-h)}.
\ee
In \eqref{Def:AdSHarm}, we have also introduced the projection operator $D_{Z_0}:= D_{Z_0^A}$ defined in \eqref{Def:DAop}, 
which executes the tensor index contractions.
Combining \eqref{Def:AdSHarm0}, \eqref{Def:Id-Pi} and \eqref{Def:AdSHarm}, we can directly relate bulk to bulk and bulk to boundary propagators through the following ``cutting identity''\footnote{Here we refrain from using the terminology of closely related ``split representation'' to avoid confusion, as discussed in Section \ref{Section:Decomposition}, split representation of bulk to bulk propagator involves boundary to boundary propagators with lower spins.}:
\be\label{bulk-boundary relation}
 \Pi_{\Del, J}(X, \tX; W, \tW) = \int^{+\infty}_{-\infty} d\nu \int_\partial dP_0 \frac{\nu^2}{\pi J! (h-1)_J} \frac{1}{\nu^2+(\Del-h)^2}  \Pi_{h+iv, J}(X, P_0; W, D_{Z_0}) \Pi_{h-iv, J}(\tX, P_0; \tilde{W}, Z_0).
 \ee
We will refer to the complex integration parameter $\nu$ as the ``spectral parameter''.
\paragraph{}
Given the relation \eqref{bulk-boundary relation}, we can now use it to rewrite the bulk to bulk propagator entering \eqref{Def:ScalarGWD1}.
More explicitly as in \eqref{Def:spinJPibb-components}, we can extract the STT tensor structures from \eqref{bulk-boundary relation} using the projection operator $K_A$:
\be\label{Def:Spin l-bulk-boundary comp}
\frac{1}{J !(\frac{d-1}{2})_{J}}K_{A_1\dots A_J}  \Pi_{\Del, J}(X, P; W, Z) = \cC_{\Del, J}\frac{(2 X\cdot \rC)_{\{A_1\dots A_J\}}}{(-2P\cdot X)^{\tau}} 
\ee
where $(2X\cdot \rC)_A = 2 X^B \rC_{ BA} = -2\rC_{AB} X^B$ satisfies both transverse $(2X\cdot \rC)_A X^A =0$ and traceless $\eta_{AB} (2X\cdot \rC)^A (2X\cdot \rC)^B = 0$ properties.
Moreover $(2X\cdot \rC)_B G^B_A = (2X\cdot \rC)_A$ simplifies the resultant expression.
\paragraph{}
Effectively upon the substitution, we have cut the four point geodesic Witten diagram into a pair of three point ones,
and we call them ``three point geodesic Witten diagrams'' or ``three point GWDs'', see Figure \ref{Fig:3pt-geostraight}. 
We can now explicitly consider the general interaction vertex at $X(\lam)$ (or $\tX(\lam')$), which includes two massive scalar fields $\Phi_{1,2}(X)$ and a rank-$J$ massive STT tensor field $\Xi_J(X)$, 
corresponding to the holographic duals of the CFT operators $\cO_{\Del_{1,2}}(P_{1,2})$ and $\cO_{h+i\nu, J}(P_0, Z_0)$:
\be\label{Scalar-3pt-Int-vertex-J}
g_{\Phi_1\Phi_2\Xi_J}\int_{X=X(\lambda)} dX \nabla^{C_1}\dots \nabla^{C_r}\Phi_1(X) \nabla^{C_{r+1}} \dots \nabla^{{C_J}} \Phi_2(X) \Xi(X)_{C_1\dots C_J}, 
\ee
where $r=1, \dots, J$ encode all possible permutations of covariant derivatives and $g_{\Phi_1\Phi_2\Xi_J}$ is the coupling constant.
In contrast with the usual three point Witten diagram, where the interaction point $X$ is integrated over the entire AdS$_{d+1}$ space $X^2=-1$, 
here we restrict the interaction point only along the geodesic  $\gamma_{12}: X=X(\lam)$. Such that when we move the covariant derivatives using integration by parts and apply equation of motion, 
we need to carefully treat the boundary terms, this has interesting effect when we consider geodesic Witten diagrams involving external spinning fields.
\paragraph{}
If we now perform the integration along $\gamma_{12}$ first, the three point vertex \eqref{Scalar-3pt-Int-vertex-J} generates the following integral:
\bea\label{3pt-ScalarGWD-J}
&&\int_{\gamma_{12}} (K\cdot\nabla)^r\frac{\cC_{\Del_1}}{(-2P_1\cdot X)^{\Del_1}} (K\cdot\nabla)^{J-r}  \frac{\cC_{\Del_2}}{(-2P_2\cdot X)^{\Del_2}} 
  \left[  \cC_{h+i\nu, J}  \frac{(2 X\cdot C_0\cdot W)^{J}}{(-2P_0\cdot X)^{h+i\nu+J}}\right]\nn\\
 && =\cC_{\Del_1}\cC_{\Del_2} \cC_{h+i\nu, J}(\Del_1)_r(\Del_2)_{J-r}      \int_{\gamma_{12}}  \frac{ (2P_1\cdot G \cdot K)^r (2P_2\cdot G \cdot K)^{J-r}}{(-2P_1\cdot X)^{\Del_1+r} (-2P_2\cdot X)^{\Del_2+J-r} }  
   \left[\frac{ (2 X\cdot C_0\cdot W)^J}{(-2P_0\cdot X)^{h+i\nu+J}}\right]\nn\\
  &&=  \cC_{\Del_1}\cC_{\Del_2} \cC_{h+i\nu, J}(\Del_1)_r(\Del_2)_{J-r}    \int^{\infty}_{-\infty} d\lambda\frac{   (-1)^{J-r}  \left(\frac{dX(\lam)}{d\lambda}\cdot K\right)^J}{  (-2P_1\cdot X(\lambda))^{\Del_1} (-2P_2\cdot X(\lam))^{\Del_2}}
  \frac{ (2X(\lam)\cdot C_0\cdot W)^{J}}{(-2P_0\cdot X(\lam))^{h+i\nu+J}}\nn\\
  &&=  \cC_{\Del_1}\cC_{\Del_2} \cC_{h+i\nu, J}(\Del_1)_r(\Del_2)_{J-r} \int^{\infty}_{-\infty} d\lam \frac{ (-1)^{J-r} J!\left(\frac{d-1}{2}\right)_J [2 {\rm V}_{0,12}]^{J} }{(-2P_1\cdot X(\lambda))^{\Del_1} (-2P_2\cdot X(\lam))^{\Del_2}  (-2P_0\cdot X(\lam))^{h+i\nu+J} }\nn\\
&&= 2^{J}(-1)^{J-r} J!\left(\frac{d-1}{2}\right)_J  (\Del_1)_r(\Del_2)_{J-r}   \cC_{\Del_1}\cC_{\Del_2} \cC_{h+i\nu, J}   \beta_{ \Del_{12}, h+i\nu+J } 
\begin{bmatrix}
				\Delta _1 &\Delta _2&  h+i\nu       \\[0.05em]
				0 & 0           & J \\[0.05em]
				0         &     0     & 0  
			\end{bmatrix}.		
\eea
The lengthy calculation presented above requires some explanations.
In the second line of \eqref{3pt-ScalarGWD-J}, we have used the identity:
\be
(W\cdot \nabla)^{J-l}\frac{(2X\cdot C_0\cdot W)^{l}}{(-2P_0\cdot X)^{\Del+l}} = (\Del+l)_{J-l} \frac{(2P_0\cdot W)^{J-l}(2X\cdot C_0\cdot W)^{l}}{(-2P_0\cdot X)^{\Del+J}} 
\label{Deriv-Id}. 
\ee
Restricting  along the geodesic $\gamma_{12}$, we also have the following relations in the third line:
\be\label{Pullback-dX}
(2P_1\cdot G)_A = (-2P_1 \cdot X(\lam))\frac{d X_A(\lam)}{d \lam}, \quad (2 P_2\cdot G)_A = -(-2P_2 \cdot X(\lam))\frac{d X_A(\lam)}{d \lam}, 
\ee
which yield the product of $\frac{dX(\lam)}{d\lam}$ and $\frac{d\tX(\lam')}{d\lam'}$ appearing in \eqref{Def:SpinJ-prop}.
Finally in the last two lines, we introduced the independent tensor basis for three point functions defined in \eqref{boxxx}   and \eqref{Def:Vijk}, 
and we have performed the integral using the result in Appendix \ref{App:Integral}. In particular, the $\nu$-dependent pre-factor is:
\be\label{beta-nu-J}
\beta_{\Del_{12}, h+i\nu+J} = \frac{\Gamma\left(\frac{h+i\nu+J+ \Del_{12}}{2}\right) \Gamma\left(\frac{h+i\nu+J-\Del_{12}}{2}\right)}{2 \Gamma(h+i\nu+J)}.
\ee 
We can also consider analogous three point vertex to \eqref{Scalar-3pt-Int-vertex-J} along the geodesic  $\gamma_{34} : \tX =\tX(\lam)$ for the holographic duals of 
$\cO_{\Del_{3,4}}(P_{3,4})$ and $\cO_{\frac{d}{2}-i\nu, J}(P_0, Z_0)$, and obtain the same tensor structure as in \eqref{3pt-ScalarGWD-J} with trivial substitution $(\Del_1, \Del_2, h+i\nu)\to (\Del_3, \Del_4, h-i\nu)$.
\paragraph{}
Gluing together the pair of resultant geodesic three point Witten diagrams for $\cO_{\Del_{1,2}}(P_{1,2}), \cO_{\frac{d}{2}-i\nu, J}(P_0, Z_0)$ and $\cO_{\Del_{3,4}}(P_{3,4}), \cO_{\frac{d}{2}-i\nu, J}(P_0, Z_0)$ given in \eqref{3pt-ScalarGWD-J},
by contracting their indices and integrating their common boundary point $P_0$, 
we obtained an integral representation of four point scalar geodesic Witten diagram $\cW_{\Del, l}(P_i)$:
\bea
&&\cS_{\Del_{1,2,3,4}}^{J, r, r'}\int dP_0 \int^{\infty}_{-\infty} d\nu \frac{\nu^2  \cC_{h+i\nu, J} \cC_{h-i\nu, J} \beta_{\Del_{12},h+i\nu+J} \beta_{\Del_{34},h-i\nu+J}
   }{\nu^2+(\Del-h)^2} 
\begin{bmatrix}
				\Delta _1 &\Delta _2&  h+i\nu    \\[0.05em]
				0 & 0           & J \\[0.05em]
				0         &     0     & 0  
			\end{bmatrix}
\cdot			
\begin{bmatrix}
				\Delta _3 &\Delta _4&  h-i\nu  \\[0.05em]
				0 & 0           & J \\[0.05em]
				0         &     0     & 0  
			\end{bmatrix}
 \nn\\
&& = \frac{ \cS_{\Del_{1,2,3,4}}^{J, r, r'} }{2^4\pi^d} \int dP_0 \int^{\infty}_{-\infty} d\nu  {\mathcal B}_{d, J}(\nu)   \frac{ \cK_{\Del_{12}, \Del_{34};J}\left(h+i\nu, h-i\nu\right) }{\nu^2+(\Del-h)^2  }
\begin{bmatrix}
				\Delta _1 &\Delta _2&  h+i\nu    \\[0.05em]
				0 & 0           & J \\[0.05em]
				0         &     0     & 0  
			\end{bmatrix}	
\cdot	
\begin{bmatrix}
				\Delta _3 &\Delta _4&  h-i\nu  \\[0.05em]
				0 & 0           & J \\[0.05em]
				0         &     0     & 0  
			\end{bmatrix}.
\nn\\
\label{Scalar-GWD-Split}
\eea
Here the overall constant is given by:
\be
\cS_{\Del_{1,2,3,4}}^{J, r, r'} = \cC_{\Del_1} \cC_{\Del_2} \cC_{\Del_3} \cC_{\Del_4} \left[2^J J! \left(\frac{d-1}{2}\right)_J\right]^2 (\Del_1)_r (\Del_2)_{J-r} (\Del_3)_{r'} (\Del_4)_{J-r'}.
\ee
The dot  ``$\cdot$''  product between the two box tensor basis for the three point functions indicates that we have replaced $Z_0$ by $D_{Z_0}$ in the first term as in \eqref{bulk-boundary relation} to perform the index contractions. We have also defined the following short hand notations and composite functions:
\bea
&&\Gamma(x\pm b) = \Gamma(x+b)\Gamma(x-b), \quad (x\pm b)_J = (x+b)_J (x-b)_J, \\
&&
{\mathcal B}_{h, J}(\nu) = \frac{1}{\left(h-1\pm i\nu\right)_J  \Gamma(\pm i\nu) },\quad
\cK_{\Del_{12}, \Del_{34}; J}(x,y) = \Gamma\left(\frac{x+J\pm \Del_{12}}{2}\right)  \Gamma\left(\frac{y+J\pm \Del_{34}}{2}\right)
.\nn\\
\eea
\paragraph{}
We can also deduce an analogous integral representation for the conformal partial wave $W_{\Del, J}(P_i)$, which also involves the so-called ``shadow operator'' $\tilde{\cO}_{\tDel, J}(P_0, Z_0)$ of the exchanged operator $\cO_{\Del, J}(P_0, Z_0)$ \cite{DO-2011} \cite{DSD-Shadow}, carrying the scaling dimension $\tDel = d-\Del$ and the same spin $J$.  
Our starting point is the equation (3.25) of \cite{DO-2011}, which relates the linear combination of the conformal block $G_{\cO_{\Del, J}}(u, v)$ and its shadow $G_{\tilde{\cO}_{\tDel, J}}(u, v)$, with an integral containing a pair of three point functions involving $\cO_{\Del, J}$ and $\tilde{\cO}_{\tDel, J}$. 
By multiplying the appropriate pre-factors as in \eqref{Def:ScalarCPW}, we can deduce the following equation for the conformal partial wave and its shadow
\footnote{We can verify by direct  computation that the embedding space building blocks  $\rV_{0,12}$ and $\rV_{0,34}$ can be projected into physical space as
$\rV_{0, 12} = \frac{x_{01}^2 x_{02}^2}{x_{12}^2} X_\nu z_0^\nu$ and  $\rV_{0, 34} = -\frac{x_{03}^2 x_{04}^2}{x_{34}^2} \bar{X}_\nu z_0^\nu$, where $X^\mu$ and $\bar{X}^{\mu}$ are the vectors defined in equations (3.6) of \cite{DO-2011}. We can then make identifications between the three point functions in physical space and the embedding space box tensor basis as in \eqref{Key-Relation1}.}
:
\bea\label{Key-Relation1}
&&(-1)^J\frac{  \cK_{\Del_{12}, \Del_{34}, J}\left(h+i\nu, h-i\nu\right) }{\pi^{\frac{d}{2}}} \int dP_0 \begin{bmatrix}
				\Delta _1 &\Delta _2&  h+i\nu    \\[0.05em]
				0 & 0           & J \\[0.05em]
				0         &     0     & 0  
			\end{bmatrix}	
\cdot	
\begin{bmatrix}
				\Delta _3 &\Delta _4&  h-i\nu \\[0.05em]
				0 & 0           & J \\[0.05em]
				0         &     0     & 0  
			\end{bmatrix}
			\nn\\
&& = \frac{1}{2^J c_J}\left( \frac{ \cK_{\Del_{12}, \Del_{34}, J}\left(h+i\nu, h+i\nu\right)}{k_{h-i\nu, J}} W_{\cO_{h+i\nu, J}}(P_i) + 
\frac{ \cK_{\Del_{12}, \Del_{34}, J}\left(h-i\nu, h-i\nu\right) }{k_{h+i\nu, J}} W_{\tilde{\cO}_{h-i\nu, J}}(P_i) 
\right),\nn\\
\eea
where we have set $\Del =h+i\nu$ and defined:
\be
c_J= \frac{\left(\frac{d-2}{2}\right)_J}{(d-2)_J}\quad k_{\Del, J} = \frac{\Gamma(\Del-1)\Gamma(d-\Del+J)}{\Gamma(\Del-1+J)\Gamma(\Del-h)}.
\ee
Now to revert the relation \eqref{Key-Relation1} and extract $W_{\cO_{\Del, J}}(P_i)$, we first multiply both sides with $\frac{\cB_{h, J}(\nu)}{\nu^2+(h-\Del)^2}$ and integrate over $\nu$,
clearly the LHS is now proportional to \eqref{Scalar-GWD-Split}, while from RHS we obtained the following integral:
\be\label{Key-Relation2}
\frac{1}{2^J c_J}\int^{\infty}_{-\infty} d\nu \frac{1}{\nu^2+(h-\Del)^2}\left( f_{\Del_{12},\Del_{34}, J}\left(h+i\nu\right)W_{\cO_{h+i\nu, J}}(P_i) + 
 f_{\Del_{12},\Del_{34}, J}\left(h-i\nu\right)  W_{\tilde{\cO}_{h-i\nu, J}}(P_i) \right),
\ee
where
\be
 f_{\Del_{12},\Del_{34}, J}\left(h\pm i\nu\right) = \frac{\cK_{\Del_{12}, \Del_{34}, J}\left(h\pm i\nu, h\pm i\nu\right)}{\Gamma(\pm i\nu)\Gamma(h\pm i\nu+ J)(h\pm i\nu-1)_J}.
\ee
In the first integrand of \eqref{Key-Relation2}, since $W_{\cO_{h+i\nu, J}}(P_i) \to 0$ as ${\rm Im}(\nu) \to -\infty$, 
assuming $\Del  > h$, we close the contour in the lower half plane to  pick up the residue at $\nu = -i(\Del-h)$, 
similarly for the second integrand of \eqref{Key-Relation2}, we can close the contour in the upper half plane to pick up the residue at $\nu = +i(\Del-h)$.
Moreover one can check that provided 
\be
J+h > |\Del_{12}|, |\Del_{34}|
\ee
is satisfied, the factor $ f_{\Del_{12},\Del_{34}, J}\left(h\pm i\nu\right)$ does not contain any additional poles in the lower/upper half plane\footnote{This is obviously satisfied by identical scalars $\Del_{12} = \Del_{34} =0$\,.}. 
Collecting all the factors we see that the conformal partial wave then takes the following integral representation\footnote{Special case of this expression for identical scalars was also obtained in \cite{Sleight-Thesis}.}:
\be\label{Scalar-CPW-integral}
W_{\Del, J}(P_i) = 
F_{\Del_{12}, \Del_{34}, J}(\Del)
\int dP_0 \int^{\infty}_{-\infty} d\nu\,   {\mathcal B}_{h, J}(\nu)  \frac{\cK_{\Del_{12}, \Del_{34}, J}\left(h+i\nu, h-i\nu\right)}{\nu^2+(\Del-h)^2}
\begin{bmatrix}
				\Delta _1 &\Delta _2&  h+i\nu    \\[0.05em]
				0 & 0           & J \\[0.05em]
				0         &     0     & 0  
			\end{bmatrix}	
\cdot	
\begin{bmatrix}
				\Delta _3 &\Delta _4&  h-i\nu \\[0.05em]
				0 & 0           & J \\[0.05em]
				0         &     0     & 0  
			\end{bmatrix}
\ee
where $F_{\Del_{12}, \Del_{34}, J}(\Del) = \frac{(-2)^J (\Del-h)c_J}{2\pi^{h+1} f_{\Del_{12}, \Del_{34},J}(\Del)}$.
We see that up to overall constants the integral representation of four point scalar geodesic Witten diagram \eqref{Scalar-GWD-Split} 
precisely matches with the integral representation of the scalar conformal partial wave \eqref{Scalar-CPW-integral}.
This provides an alternative proof of the results in \cite{ScalarGWD}. 
In particular, we have done so by using the relation \eqref{bulk-boundary relation} to build the scalar four point geodesic Witten diagram using three point ones,
through the integration with the corresponding measure as in \eqref{Scalar-GWD-Split}.
This is completely analogous to how we construct the conformal blocks using three point functions. 
\paragraph{}
Closely related computation has been done in \cite{SpinningAdS} and \cite{SpinningAdS2}, which builds the four point Witten diagrams from the three point ones\footnote{More comprehensive analysis can also be found in \cite{Sleight-Thesis}.},
it is somewhat expected that the three point Witten and geodesic Witten diagrams for scalar-scalar-spin-$J$ exchange are both proportional to the same tensor structure, as there is only single one available. The crucial difference here however 
is the different $\nu$-dependent pre-factors generated through integration over entire AdS$_{d+1}$ and only along geodesics. 
The pre-factor for three point Witten diagrams, upon integrating with the same $\frac{1}{\nu^2+(\Del-h)^2}$ typically yields conformal blocks for operator $\cO_{\Del, J}$ plus infinite towers of double trace operator $\cO_{\Del_m^{(12)},\,l}$ and $\cO_{\Del_n^{(34)},\,l}$ where $0\leq l<J$ and the dimensions $\Delta^{(12)}_{m,\,l}$ and $\Delta^{(34)}_{m,\,l}$ are
defined in \eqref{Def:dim-double}.
While the corresponding pre-factor for three point {\it geodesic} Witten diagram \eqref{beta-nu-J} does not contain these infinite double trace operator poles, 
such that upon integration \eqref{Scalar-GWD-Split} we precisely only have conformal block for $\cO_{\Del, J}$ exchange.
In section \ref{Section:Decomposition}, we will start from the so-called ``split representation'' of 
the four point Witten diagram to recover their decompositions into four geodesic Witten diagrams for single and double trace operator exchanges with arbitrary spins.
We will see that the three point geodesic Witten diagrams play the role of building block for various four point geodesic Witten diagrams. 

\section{Spinning Three Point Functions and Conformal Blocks}\label{Sec:SpinningCFT}
\paragraph{}
Having demonstrated how the integral representation of scalar conformal partial waves can be directly realized through cutting up the four point scalar GWDs, 
and identify the resultant three point GWDs with the three point correlation functions,
this procedure becomes even more useful when systematically constructing the holographic dual configuration for conformal partial waves/conformal blocks for external operators carrying arbitrary integer spins.
Here we will restrict ourselves here to only the exchange of symmetric traceless field, as in the case of scalar conformal partial waves we just reviewed.
Even though there can be additional exchange channels involving mixed tensor fields (see e. g. \cite{MixedTensor1, MixedTensor2, SpinRadial1} ), we leave the detailed holographic analysis to the future work.  
\paragraph{}
Let us begin by reviewing CFT side of the story, this was done throughly in \cite{SpinningBlock0, SpinningBlock1}. 
The external primary operators with scaling dimension $\Del_i$ and spin $l_i$ are labeled as $\cO_{\Del_i, l_i}(P_i, Z_i)$, where $P_i$ is the position in the embedding space as before and $Z_i$ is the auxiliary polarization vector. Such that $\cO_{\Del_i, l_i}(P_i, Z_i)$ is a homogenous polynomial in $Z_i$ of degree $l_i$, and we can recover the STT tensor field in embedding space through differential operator $D_{Z_i}$ defined in \eqref{Def:DAop}.
We again collect the relevant details about the embedding space representatives for $d$-dimensional tensors in the Appendix \ref{Appendix:Embed}.
\paragraph{}
The three point correlation functions involving $\{ \cO_{\Del_i, l_i}(P_i, Z_i)  \}$ are crucial building blocks for higher point correlation functions, 
their form can be completely fixed by conformal symmetries manifest in the embedding space, which lead to the classifications in \cite{SpinningBlock0, SpinningBlock1}:
\be\label{Spinning3pt}
<\cO_{\Del_1, l_1}(P_1, Z_1)\cO_{\Del_2, l_2}(P_2, Z_2)\cO_{\Del_3, l_3}(P_3, Z_3) > = \sum_{n_{12}, n_{13}, n_{23}\ge 0}\lambda_{n_{12}, n_{13}, n_{23}}
\begin{bmatrix}
				\Delta _1 &\Delta _2&  \Del_3   \\[0.05em]
				l_1 & l_2           & l_3 \\[0.05em]
				n_{23}         &     n_{13}     & n_{12}  
			\end{bmatrix}
+\cO(Z_i^2, Z_i\cdot P_i).
 \ee
Here $\lambda_{n_{12}, n_{13}, n_{23}}$ are theory dependent constant expansion coefficients, and in addition to the integer spins $\{l_i\}$, we have also introduced triplet of non-negative integers $\{n_{12}, n_{13}, n_{23}\}$ satisfying the following constraint:
\be\label{Def:m123}
m_1 = l_1-n_{12}-n_{13} \ge 0, \quad  m_2 = l_2-n_{12}-n_{23} \ge 0 \quad  m_3 = l_3-n_{13}-n_{23} \ge 0.
\ee
The elementary structures of the three point correlation function, which we shall call {\it ``box tensor basis''} are then given by:
 \be\label{boxxx}
	 \begin{bmatrix}
	 	\Delta _1 &\Delta _2&  \Delta _3        \\[0.05em]
	l_1 & l_2           & l_3\\[0.05em]
	n_{23}         &     n_{13}      &n_{12}  
	\end{bmatrix}=\frac{{\rV}_{1,23}^{m_1}{\rV}_{2,31}^{m_2}{\rV}_{3,12}^{m_3}{\rH}_{12}^{n_{12}}{\rH}_{13}^{n_{13}}{\rH}_{23}^{n_{23}}}{\left(P_{12}\right){}^{\frac{1}{2}\left(\tau _1+\tau _2-\tau _3\right)}\left(P_{13}\right){}^{\frac{1}{2}\left(\tau _1+\tau _3-\tau _2\right)}\left(P_{23}\right){}^{\frac{1}{2}\left(\tau _2+\tau _3-\tau _1\right)}}. 
	\ee
Here we have defined the six linearly independent tensor basis for three operators with integer spins:
\bea
\rH_{ij} &=& -2\left[(Z_i\cdot Z_j)(P_i\cdot P_j)-(Z_i\cdot P_j)(Z_j\cdot P_i)\right]=-{\rm Tr}(\rC_i\cdot \rC_j) \label{Def:Hij},\\
\rV_{i, jk} &=& \frac{\left(P_j \cdot Z_i\right)\left(P_i\cdot P_k\right)-\left(P_j\cdot P_i\right)\left(Z_i\cdot P_k\right)}{\left(P_j\cdot P_k\right)}=\frac{(P_j\cdot \rC_i \cdot P_k)}{( P_j\cdot P_k)}, \quad i, j, k=1,2,3 \label{Def:Vijk}.
\eea 	
Notice that $\rH_{jk}$ is symmetric and $\rV_{i, jk}$ is anti-symmetric under the exchange of $j$ and $k$ indices, such that we only have altogether six independent basis.
They combine to form transverse polynomial of degree $l_i$ in each $Z_i$ (also each $P_i$) in the numerator of \eqref{boxxx}, 
the tensorial structure of \eqref{boxxx} is revealed through the action of $D_{Z_i}$ operators.
The number of the set of non-negative integers satisfying \eqref{Def:m123} is the possible elementary structures listed in \eqref{boxxx}, 
for $l_3 \ge l_2 \ge l_1$ and $p ={\rm max}(0, l_1+l_2-l_3)$, it is given by:
\be\label{N123}
N(l_1, l_2, l_3) = \frac{(l_1+1)(l_1+2)(3l_2-l_1+3)}{6} - \frac{p(p+2)(2p+5)}{24}-\frac{1-(-1)^p}{16}.
\ee
The remaining terms labeled $\cO(Z_i^2, Z_i\cdot P_i)$ arise from $(\rC_i\cdot \rC_i)$, $(\rC_i\cdot P_i)$ or $(\rC_i\cdot Z_i)$ types of contractions which are not independent and can be determined
when taking into account of light cone and transversality conditions.
\paragraph{}
Another very useful basis for expressing the structures of three point functions involve the following differential operators:
\bea
&&\rD_{11} = \left((P_1\cdot P_2) Z_1^A-(Z_1\cdot P_2)P_1^A\right)\frac{\pa}{\pa P_2^A} + \left((P_1\cdot Z_2)Z_1^A-(Z_1\cdot Z_2) P_1^A\right)\frac{\pa}{\pa Z_2^A}\label{Def:D11},\\
&&\rD_{12} = \left((P_1\cdot P_2) Z_1^A-(Z_1\cdot P_2)P_1^A\right)\frac{\pa}{\pa P_1^A} + \left((P_2\cdot Z_1)Z_1^A\right)\frac{\pa}{\pa Z_1^A}\label{Def:D12},\\
&&\rD_{22} = \left((P_1\cdot P_2) Z_2^A-(Z_2\cdot P_1)P_2^A\right)\frac{\pa}{\pa P_1^A} + \left((P_2\cdot Z_1)Z_2^A-(Z_1\cdot Z_2) P_2^A\right)\frac{\pa}{\pa Z_1^A}\label{Def:D22},\\
&&\rD_{21} = \left((P_1\cdot P_2) Z_2^A-(Z_2\cdot P_1)P_2^A\right)\frac{\pa}{\pa P_2^A} + \left((P_1\cdot Z_2)Z_2^A\right)\frac{\pa}{\pa Z_2^A}\label{Def:D21},
\eea
and they only have the following non-vanishing commutators:
\bea
&&[\rD_{11}, \rD_{22}] =\frac{\rH_{12}}{2}\left(Z_1\cdot \frac{\pa}{\pa Z_1}-Z_2\cdot\frac{\pa}{\pa Z_2}+P_1\cdot \frac{\pa}{\pa P_1}-P_2\cdot\frac{\pa}{\pa P_2}\right),\label{D-commutator1}\\
&&[\rD_{12}, \rD_{21}] =\frac{\rH_{12}}{2}\left(Z_1\cdot \frac{\pa}{\pa Z_1}-Z_2\cdot\frac{\pa}{\pa Z_2}-P_1\cdot \frac{\pa}{\pa P_1}+P_2\cdot\frac{\pa}{\pa P_2}\right), \label{D-commutator2}
\eea
while all other commutators vanish, including $[\rD_{ij}, \rH_{12}] = 0$.
We shall express such differential basis using curly brackets, and they are defined through the following relations:
\bea \label{differBasis}
 \begin{Bmatrix}  
	\Delta _1 &\Delta _2&  \Delta _3        \\[0.01em]
	l_1 & l_2           & l_3\\[0.01em]
	n_{23}         &     n_{13}      &n_{12}  
	\end{Bmatrix}&=&{\rH}_{12}^{n_{12}}\rD_{12}^{n_{13}}\rD_{21}^{n_{23}}\rD_{11}^{m_{1}}\rD_{22}^{m_{2}} \Sigma^{l_1+n_{23}-n_{13}, l_2+n_{13}-n_{23}}	
	\begin{bmatrix}
				\Delta _1 &\Delta _2 &  \Delta _3        \\[0.1em]
				0 & 0           & l_3\\[0.1em]
				0         &     0      & 0
			\end{bmatrix},\nn\\
			&=&{\rH}_{12}^{n_{12}}\rD_{12}^{n_{13}}\rD_{21}^{n_{23}}\rD_{11}^{m_{1}}\rD_{22}^{m_{2}} 	
	\begin{bmatrix}
				\tilde{\tau} _1 &\tilde{\tau} _2 &  \Delta _3        \\[0.1em]
				0 & 0           & l_3\\[0.1em]
				0         &     0      & 0
			\end{bmatrix},			
\eea
where the shift operators $\Sigma^{a, b}$ which shifts the scaling dimensions $(\Del_1, \Del_2)$ to $(\Del_1+a, \Del_2+b)$, 
such that $\tilde{\tau}_1 = \tau_1+(n_{23}-n_{13})$ and  $\tilde{\tau}_2 = \tau_2+(n_{13}-n_{23})$.
Notice that for given integer spins $\{l_1, l_2, l_3\}$, \eqref{differBasis} are also labeled by triplet of non-negative integers $\{n_{12}, n_{13}, n_{23}\}$ satisfying \eqref{Def:m123},
we therefore have equal number $N(l_1, l_2, l_3)$ of differential basis \eqref{differBasis} as in the original box basis \eqref{boxxx}, and they are related by linear transformation with constant coefficients.
\paragraph{}
In contrast with the box basis \eqref{boxxx}, where we can cyclicly permute the three primary operators involved, 
in the differential basis we break this cyclicity such that the differential operators \eqref{Def:D11}-\eqref{Def:D21} only act on $(P_{1,2}, Z_{1,2})$,
moreover the remaining box tensor structure in the RHS of \eqref{differBasis} is precisely the one arising in the integral representation of conformal partial wave \eqref{Scalar-CPW-integral} 
after identifying $(\Del_3, l_3)$ with $(h+i\nu, J)$.
We can therefore regard the remaining primary operator as the internal exchange operator when constructing the four point correlation function, 
this allows us to relate four point correlation functions for operators with spins:
\be\label{Spinning 4pt function}
<\cO_{\Del_1, l_1}(P_1, Z_1)\cO_{\Del_2, l_2}(P_2, Z_2)\cO_{\Del_3, l_3}(P_3, Z_3) \cO_{\Del_4, l_4}(P_4, Z_4) >,
\ee
with the scalar ones.
More explicitly, unlike scalar case \eqref{Def:ScalarCPW} whose conformal partial wave for a given exchanged operator $\cO_{\Del, J}$  
can be packaged into a single scalar function of cross-ratios;
the conformal partial wave for \eqref{Spinning 4pt function} for a given exchange operator 
consists of multiple terms each with independent tensor structures. 
When restricting to only the exchange of symmetric traceless operators, we can construct it by fusing the differential basis for a pair of three point correlation functions involving primary operators $\cO_{\Del_{1,2}, l_{1,2}}, \cO_{\Del, J}$ and  $\cO_{\Del_{3,4}, l_{3,4}}, \tilde{\cO}_{d-\Del, J}$
and the resultant conformal partial wave schematically contains the following tensor structures: 
\be\label{Def:SpinningBasis}
W_{\cO_{\Del, J}}^{\{n_{10}, n_{20}, n_{12}\}; \{n_{30}, n_{40}, n_{34}\}}(P_i, Z_i):= \cD_{\rm Left}^{n_{10}, n_{20}, n_{12}} \cD_{\rm Right}^{n_{30}, n_{40}, n_{34}} W_{\cO_{\Delta, J}} (P_i). 
\ee
Here the composite operators are given by:
\bea
\cD_{\rm Left}^{n_{10}, n_{20}, n_{12}} &=& \rH_{12}^{n_{12}}\rD_{12}^{n_{10}} \rD_{21}^{n_{20}}\rD_{11}^{m_1} \rD_{22}^{m_2}\Sigma^{l_1+n_{20}-n_{10}, l_2-n_{20}+n_{10}},\label{Def:DLeft}\\
\cD_{\rm Right}^{n_{30}, n_{40}, n_{34}} &=& \rH_{34}^{n_{34}}\rD_{34}^{n_{30}} \rD_{43}^{n_{40}}\rD_{33}^{m_3} \rD_{44}^{m_4}\Sigma^{l_3+n_{40}-n_{30}, l_4-n_{40}+n_{30}}. \label{Def:DRight}
\eea
They are now labeled by two sets of triplet of integers $\{n_{10}, n_{20}, n_{12}\}$ and $\{n_{30}, n_{40}, n_{34}\}$ satisfying \eqref{Def:m123}, 
and we have denoted the exchanged operator as $\cO_{\Del_0, l_0}\equiv \cO_{\Del, J}$. 
Summing over all $N(l_1, l_2, l_0)\times N(l_3, l_4, l_0)$ possible tensor structures listed in \eqref{Def:SpinningBasis}, we can express the resultant conformal partial wave for $\cO_{\Del, J}$ as: 
\be
\sum_{\{n_{10}, n_{20}, n_{12}\}\\  \{n_{30}, n_{40}, n_{34}\}}  W_{\cO_{\Del, J}}^{\{n_{10}, n_{20}, n_{12}\}; \{n_{30}, n_{40}, n_{34}\}}(P_i, Z_i)
= \left(\frac{P_{24}}{P_{14}}\right)^{\frac{\tau_1-\tau_2}{2}} \left(\frac{P_{14}}{P_{13}}\right)^{\frac{\tau_3-\tau_4}{2}}\frac{\sum_{k} f_k(u, v) Q^{(k)}(P_i; Z_i)}{(P_{12})^{\frac{\tau_1+\tau_2}{2}} (P_{34})^{\frac{\tau_3+\tau_4}{2}}}.
\ee
Here $Q^{(k)}({P_i, Z_i})$ are transverse polynomials of degree $l_i$ in $Z_i$ and can be built from $\rH_{12}$, $\rH_{34}$ and $\rV_{i, jk}$ given in \eqref{Def:Hij} and \eqref{Def:Vijk}, now with $i, j, k = 1,2,3,4$. $f_k(u, v)$ denote the functions of purely cross ratio $(u, v)$ which can be obtained by mechanical differentiations involving $\rD_{ij}$ operators, 
and they consist of derivatives of the scalar conformal block for the same exchange operator $G_{\cO_{\Del, J}}(u, v)$ with respect to cross ratios $(u, v)$.
It is interesting to note that all the differential operators $\rD_{ij}$ \eqref{differBasis} only act on the external position and polarization vectors $(P_i, Z_i)$, 
we can thus readily obtain conformal partial waves for spinning primary operators in terms of the scalar ones.
\paragraph{}
We can also easily deduce the integral representation for the spinning conformal blocks from the one for scalar conformal block \cite{DSD-Shadow}. 
This amounts to simply replacing the box tensor structures for $\cO_{\Del_{1,2}}(P_{1,2})$ and $\cO_{h+i\nu, J}(P_0, Z_0)$ three point function appearing in \eqref{Scalar-CPW-integral} with the differential basis \eqref{differBasis}, and identify $(\Del, l_3)$ with $(h+i\nu, J)$, and similarly for the  $\cO_{\Del_{3,4}}(P_{3,4})$ and the shadow $\tilde{\cO}_{h-i\nu, J}(P_0, Z_0)$ three point function.
The result is thus:
\bea\label{SpinningCPW-Integral}
&&W_{\cO_{\Del, J}}^{\{n_{10}, n_{20}, n_{12}\}; \{n_{30}, n_{40}, n_{34}\}}(P_i, Z_i) \nn\\
&&\propto
\hat{\cD}_{\rm Left}^{n_{10}, n_{20}, n_{12}}  \hat{\cD}_{\rm Right}^{n_{30}, n_{40}, n_{34}}
\int dP_0 \int^{\infty}_{-\infty} d\nu\,   {\mathcal B}_{h, J}(\nu)  \frac{\cK_{\tilde{\tau}_{12}, \tilde{\tau}_{34}, J}\left(h+i\nu, h-i\nu\right)}{\nu^2+(\Del-h)^2}
\begin{bmatrix}
				\tilde{\tau} _1 &\tilde{\tau} _2&  h+i\nu    \\[0.05em]
				0 & 0          & J \\[0.05em]
				0        &     0     & 0  
			\end{bmatrix}	
\cdot	
\begin{bmatrix}
				\tilde{\tau} _3 &\tilde{\tau} _4&  h-i\nu \\[0.05em]
				0 & 0          & J \\[0.05em]
				0         &     0     & 0  
			\end{bmatrix},\nn\\
&&=
\int dP_0 \int^{\infty}_{-\infty} d\nu\,   {\mathcal B}_{h, J}(\nu)  \frac{\cK_{\tilde{\tau}_{12}, \tilde{\tau}_{34}, J}\left(h+i\nu, h-i\nu\right)}{\nu^2+(\Del-h)^2}
\begin{Bmatrix}
				\Del _1 &\Del _2&  h+i\nu    \\[0.05em]
				l_1 & l_2          & J \\[0.05em]
				n_{20}        &     n_{10}     & n_{12}  
			\end{Bmatrix}	
\cdot	
\begin{Bmatrix}
				\Del _3 &\Del _4&  h-i\nu \\[0.05em]
				l_3 & l_4          & J \\[0.05em]
				n_{40}         &     n_{30}     & n_{34}  
			\end{Bmatrix}.	
\eea
where $\hat{\cD}_{\rm Left}$ and $\hat{\cD}_{\rm Right}$ are defined similarly to \eqref{Def:DLeft} and $\eqref{Def:DRight}$ up the shift operators $\Sigma^{a,b}$, whose action on the scaling dimensions has been absorbed into the integral. This makes clear that we have a integral representation of a scalar conformal partial wave in the second line above with $\{\Del_i\} \to \{\tilde{\tau}_i\}$, 
followed by the action of differential operators $\hat{\cD}_{\rm Left}$ and $\hat{\cD}_{\rm Right}$.
We will see in the next section that exactly the same integral representation naturally appearing in the holographic reconstruction of the spinning conformal blocks.

\section{Spinning Conformal Partial Waves from Anti-de Sitter Space}\label{Section:AdS}
\paragraph{}
Let us begin  holographic reconstruction of spinning conformal partial waves given in terms of the basis in \eqref{Def:SpinningBasis}. 
Our strategy is simple, given the success of cutting up the four point geodesic Witten diagram into three point ones to reproduce the integral representation of scalar conformal blocks reviewed in Section \ref{Section:Review}, 
we will again first consider the geodesic three point Witten diagrams involving the holographic duals of spinning primary operators $\cO_{\Del_{1,2}, l_{1,2}}(P_{1,2}, Z_{1,2})$
and the operator $\cO_{\Del_0, l_0}(P_0, Z_0)$ in their operator product expansion. We will first prove that all possible conformally invariant three point interaction vertices, when restricting along geodesic, 
can be expressed as linear combinations of the box tensor basis given in \eqref{boxxx}, where the expansion coefficients only depend on scalar products $(P_i\cdot P_j)$, $i, j =1,2,3,4$.
Given the box tensor basis can be cast into differential tensor basis \eqref{differBasis} by linear transformations, 
moreover the composite differential operators \eqref{Def:DLeft} and \eqref{Def:DRight} commute with the integration over boundary point $P_0$,  
we can apply the same gluing procedure as in the scalar case to obtain the holographic reconstruction of the various integral representation of spinning conformal partial waves schematically given in \eqref{Def:SpinningBasis}.
\paragraph{}
Working again in the $d+2$ dimensional embedding space, let us begin by considering all possible non-vanishing Lorentz invariant contractions among three bulk to boundary propagators of scale dimensions $\Del_{1,2,0}$ and spins $l_{1,2, 0}$
with metric tensor $\eta_{AB}$ and arbitrary number of covariant derivatives $\nabla_A$. 
They can appear in the integrand for three point functions generated by all possible three point interaction vertices, whose explicit form we will discuss momentarily.
Using the equation \eqref{Deriv-Id} and $\nabla_{A} (X\cdot \rC_i)_B = G_A^{A'} G_B^{B'} \rC_{i A' B'}$, we can see that the numerator in a generic term
consists of all possible invariant contractions among $P_i^A$, $\rC_i^{ AB}$ and $X^A$;
\begin{eqnarray}
\int_{X=X(\lambda)} d\lambda ~\frac{\bbQ(\{P_i, Z_i,X\})}{(P_{1}\cdot X)^{\tau_1}(P_{2}\cdot X)^{\tau_2}(P_{0}\cdot X)^{\tau_0}}\,.
\end{eqnarray}
Restricting the bulk coordinate $X$ along the geodesic $\gamma_{12}$ given in \eqref{Def:X-geodesics}, the polynomial $\bbQ$ now only depends only $P_{1,2}$ and $Z_{1,2}$.
Moreover $\bbQ$ is invariant under the shift $Z_i \to Z_i+\alpha P_i$, as $\bbQ$ depends on $Z_i$ only through $\rC_i$ and $\rC_i$ is invariant under such a shift.
According to the discussion in \cite{SpinningBlock1} or done in more details in Appendix \ref{rewritting-ts}, 
it can be represented by using only $\rH_{ij}$ and $\rV_{ijk}$ defined in \eqref{Def:Hij} and \eqref{Def:Vijk} respectively.
Therefore three point geodesic Witten diagram with an arbitrary interaction gives a linear combination of the box tensor basis \eqref{boxxx}.
\paragraph{}
Next we would like to consider the complete three point interaction vertices involving three symmetric traceless fields in AdS$_{d+1}$ for the three point geodesic Witten diagrams,
in terms of the embedding coordinates, it can be succinctly written in the following form:
	\be\label{Def:3ptVertex-geo}
	\cV_{l_1, l_2, l_0} =\sum_{0\le \rn_r\le l_r} g_{l_1, l_2, l_0}^{\rn_1, \rn_2, \rn_0} \cJ_{l_1,l_2, l_0}^{\rn_1, \rn_2,\rn_0}(\cT^r), \quad r =1, 2, 3.
	\ee
Here $\{g_{l_1, l_2, l_0}^{\rn_1, \rn_2, \rn_0}\}$ are the theory dependent bulk coupling constants which can be eventually related the CFT OPE coefficients,
and the integers $\{\rn_1, \rn_2, \rn_0\}$ need to satisfy the conditions\footnote{Notice that while $\{\rn_1, \rn_2, \rn_0\}$ satisfy the same conditions $\{n_{10}, n_{20}, n_{12}\}$ \eqref{Def:m123}, 
as we will see from explicit computation, they are not directly identified with each other in obvious manner.}:
\be\label{n1n2n0-condition}
l_1 - \rn_2-\rn_0 \ge 0, \quad l_2 - \rn_1 -\rn_0 \ge 0, \quad l_0 - \rn_1-\rn_2 \ge 0.
\ee
While the interaction vertices along the geodesic $\gamma_{12}$ are parameterized by: 
\bea\label{Def:3ptVertex-General}
	&&\cJ_{l_1, l_2, l_0}^{\rn_1, \rn_2,\rn_0}(\cT^r) = \cY_1^{l_1-\rn_2-\rn_0} \cY_2^{l_2-\rn_0-\rn_1} \cY_3^{l_0-\rn_1-\rn_2} \cH_1^{\rn_1} \cH_2^{\rn_2} \cH^{\rn_0}_0 \cT^1(X_1, W_1) \cT^2(X_2, W_2)
	\cT^0(X_0, W_0)\mid_{X_r = X(\lam)},\nn\\
	&=&\left(\eta^{A_1 B_1} \dots \eta^{A_{\rn_0}B_{\rn_0}}\right) \left(\eta^{A_{\rn_0+1} C_1} \dots \eta^{A_{\rn_0+\rn_2}C_{\rn_2}}\right) \left(\eta^{B_{\rn_0+1} C_{\rn_2+1}} \dots \eta^{B_{\rn_0+\rn_1}C_{\rn_2+\rn_1}}\right)\nn\\
	&\times & \left[\frac{\pa}{\pa X}\right]^{(C_{\rn_1+\rn_2+1}\dots C_{l_0})}\cT^1_{\{A_1\dots A_{l_1}\}}(X) \cT^2_{\{B_1\dots B_{l_2}\}}(X)
	\left[\frac{\pa}{\pa X}\right]^{(A_{\rn_2+\rn_0+1}\dots A_{l_1})(B_{\rn_1+\rn_0+1}\dots B_{l_2})}\cT^0_{\{C_1\dots C_{l_0}\}}(X)\mid_{X=X(\lam)}, \nn\\
	\eea
	where $\cT^r_{\{A_1\dots A_{l_r}\}}(X)$ is a STT embedding space tensor field which is projected to symmetric traceless tensor field in AdS$_{d+1}$ and various differential operators are defined to be:
	\bea
	&&\cY_1= \partial_{W_1} \cdot \partial_{X_0}, \quad \cY_2= \partial_{W_2} \cdot \partial_{X_0}, \quad \cY_3= \partial_{W_0} \cdot \partial_{X_1}, \label{Def:Y-operators}\\
	&& \cH_1= \partial_{W_2} \cdot \partial_{W_0}, \quad \cH_2= \partial_{W_0} \cdot \partial_{W_1}, \quad \cH_0= \partial_{W_1} \cdot \partial_{W_2}. \label{Def:H-operators}
	\eea
Here we have almost adopted the general parameterizations found in \cite{3pt-Coupling0, 3pt-Coupling} with an essential modification on the choice of operator $\cY_1$, which is changed from $\partial_{W_1}\cdot \partial_{X_2} \to \partial_{W_1} \cdot \partial_{X_0}$,
we shall now explain the need for this modification.
Notice that in original parameterization, which integrates over the entire AdS space, 
such a change is equivalent up to equation of motion and a boundary term which we can safely discard.
However restricting along the geodesic $\gamma_{12}$, we have made an explicit choice of external legs, i.e. the curves connecting $X(\lambda)$ and $P_{1,2}$ and internal leg connecting $X(\lam)$ and $P_0$ which will be joined to form four point geodesic Witten diagram as in Section \ref{Section:Review}, such a cyclic symmetry permuting the three tensor fields is explicitly broken.
If we use the original parameterization, certain tensor structures appearing in the corresponding CFT three point function become missing.
\paragraph{}
Let us workout a simple example of spin-scalar-scalar $(l,0,0)$ case to illustrate this. First we consider the parameterization used in \cite{3pt-Coupling0, 3pt-Coupling} 
	\be\label{AdSvertex1}
	\cI_{l,0,0}^{0,0,0}=\left(\partial _{W_1}\cdot\partial _{X_2}\right){}^l\cT _1\left(X_1,W_1\right)\cT _2\left(X_2,W_2\right)\cT _0\left(X_0,W_0\right)\mid_{X_r = X},
	\ee
and when we apply this vertex to integrate over the entire AdS-space, we have:
	\bea \label{show1}
&&\int _{\rm AdS} dX \frac{\left(2P_{2}\cdot \rC_1\cdot X\right)^l}{\left(-2P_1\cdot X\right){}^{ \Delta _1+l}}\frac{1}{\left(-2P_2 \cdot X\right){}^{ \Delta _2+l}}\frac{1}{\left(-2P_0 \cdot X\right){}^{ \Delta _0}} \nn\\
&&\propto \left(P_2\cdot D_{P_1}\right)^l\mathcal{A}_3^{\Delta _1,\Delta _2, \Delta _0}\left(P_1,P_2,P_0\right)\propto [\rV_{1,20}]^{l}\mathcal{A}_{3}^{\Delta _1+l,\Delta _2, \Delta _0}\left(P_1,P_2,P_0\right).
\eea
Here $D_{P_i}^A$ is given by:
 \be\label{Def:Dpi}
D_{P_i}^{A}=Z_i^A\left(Z_i \cdot \frac{\partial }{\partial Z_i}-P_i \cdot  \frac{\partial }{\partial P_i}\right)+P_i^A\left(Z_i \cdot  \frac{\partial }{\partial P_i}\right)
\ee
and $\cA_3^{\Del_1, \Del_2, \Del_0}$ is given by the scalar integral \eqref{calc-A} in the appendix.
This vertex \eqref{AdSvertex1} precisely reproduces the only and correct corresponding tensor structure in CFT side as we expected. 
However if we use the same interaction vertex as before but now restricted along geodesic $\gamma_{12}$:
		\be\label{GWDvertex1}
	\cI_{l,0,0}^{0,0,0}=\left(\partial _{W_1}\cdot\partial _{X_2}\right){}^l\cT _1\left(X_1,W_1\right)\cT _2\left(X_2,W_2\right)\cT _0\left(X_0,W_0\right)\mid_{X_r = X(\lam)},
	\ee
we now have
		\be
		  \int _{-\infty}^{+\infty} d\lam\frac{\left(2P_2\cdot \rC_1 \cdot X(\lam)\right)^l}{\left(-2P_1\cdot X(\lam)\right){}^{ \Delta _1+l}}\frac{1}{\left(-2P_2 \cdot X(\lam)\right){}^{ \Delta _2+l}}\frac{1}{\left(-2P_0 \cdot X(\lam)\right)^{ \Delta _0}} =0
		 \ee
due to the accidental orthogonality condition $2P_2\cdot \rC_1\cdot X(\lambda)=0$ which only occurs along $\gamma_{12}$\footnote{Similar cancelation was also noted in the recent preprint \cite{Castro1}.}.	
Now if use the new parametrization given in  \eqref{Def:3ptVertex-General} instead, again we only have one type of interaction given by:
\be
\cJ_{l, 0, 0}^{0,0,0}=\left(\partial _{W_1}\cdot\partial _{X_0}\right){}^l\cT _1\left(X_1,W_1\right)\cT _2\left(X_2,W_2\right)\cT _0\left(X_0,W_0\right)\mid_{X_r = X(\lam)}.
\ee
The corresponding computation along the geodesic $\gamma_{12}$ is given by (up overall constant):
	\be
				\int _{-\infty}^{+\infty} d\lam \frac{\left(2P_0\cdot \rC_1 \cdot X(\lam)\right){}^l}{\left(-2P_1\cdot X(\lam)\right){}^{ \Delta _1+l}}\frac{1}{\left(-2P_2\cdot X(\lam)\right){}^{ \Delta _2}}\frac{1}{\left(-2P_0 \cdot X(\lam)\right){}^{ \Delta _0+l}}
				\propto  [{\rV}_{1,20}]^l \mathcal{A}_{3}^{\Delta _1+l\Delta _2\Delta _0}
	\ee
where we have used $2P_0\cdot \rC_1\cdot X (\lambda)=-e^{-\lam} \sqrt{-2P_1\cdot P_2}\rV_{1,02}$.
We have now seen that the modified parameterization instead gives the desired CFT tensor structure. 
\paragraph{}
We shall adopt the minimally modified parameterization \eqref{Def:3ptVertex-geo} in our computation of the three point geodesic Witten diagrams for symmetric traceless tensor fields. 
One important feature here is that for given $(l_1,l_2,l_0)$, the allowed range of the non-negative integers $\{\rn_1, \rn_2, \rn_0\}$ imply that 
we have the same number \eqref{N123} of independent interaction vertices as the independent box tensor structures given in \eqref{boxxx}, 
this implies that we should be able to express the resultant three point GWDs as linear combinations of these box tensor structures, echoing our general argument in the beginning of this section. Moreover as shown in \cite{3pt-Coupling}, the three point Witten diagrams produced by the original parameterization of three point vertices can also be expressed in terms of the same set of box tensor structures, this implies that we should also be able to expand the ordinary three point Witten diagrams in terms of three point GWDs. We will explicitly do so in a example that follows.
One further remark is that the we have chosen $\cY_3= \partial_{W_0} \cdot \partial_{X_1}$ in \eqref{Def:3ptVertex-geo}, the possible choice is $\cY_3 = \partial_{W_0} \cdot \partial_{X_2}$. 
But this choice is equivalent to starting with cyclically permuted three point vertices in \cite{3pt-Coupling}, then make similar modification of the differential operator to switch the partial derivative to act on $X_0$.
We believe for this other choice and the story should go through the same.

\subsection{The $(l_1,l_2,0)$ case}
\paragraph{}
Let us first consider the case with two external symmetric tensor fields with spins $l_{1,2}$ and one internal scalar field. We have the counting:
\be
l_0=0,~~l_1-\rn_{0}\geq 0, ~~l_2-\rn_{0}\geq 0,~~\rn_{1}=\rn_{2}= 0.
\ee
The corresponding interaction vertices in this case are: 
\be
			\cJ_{l_1,l_2,0}^{0,0,\rn_{0}}=(\partial_{W_1} \cdot \partial_{X_0})^{l_1-\rn_{0}} (\partial_{W_2} \cdot \partial_{X_0})^{l_2-\rn_{0}}(\partial_{W_1} \cdot \partial_{W_2})^{\rn_0} \cT^1(X_1, W_1) \cT^2(X_2, W_2)
			\cT^0(X_0, W_0)\mid_{X_r = X(\lam)}
			\ee
which yield the following integral:
\bea \label {ll0}
&&\bbC \int _{\gamma _{12}}\eta ^{A_1B_1}\dots\eta ^{A_{n_{0}}B_{n_{0}}}\frac{(2X \cdot \rC_1)_{A_1\dots A_{l_1}}}{\left(-2P_1\cdot X\right){}^{\tau _1}}\frac{(2X \cdot \rC_2)_{B_1\dots B_{l_2}}}{\left(-2P_2 \cdot X\right){}^{\tau _2}}\left(\frac{\partial }{\partial X}\right)^{A_{n_{0}+1}\dots A_{l_1}}\left(\frac{\partial }{\partial X}\right)^{B_{n_{0}+1}\dots B_{l_2}}\frac{1}{\left(-2P_0 \cdot X\right){}^{\Delta _0}}\nn \\
&&=\bbC2^{l_1+l_2-2n_0}
					(-1)^{l_1-n_0}\beta _{\tau_{12},\Delta _0}\left(\frac{\tau _{12}+\Delta _0}{2}\right)_{l_2-n_0}\left(\frac{\Del_0-\tau _{12}}{2}\right)_{l_1-n_0}\begin{bmatrix}
						\Delta _1 &\Delta _2&  \Delta _0      \\[0.05em]
						l_1 &   l_2           & 0 \\[0.05em]
						0       &     0     & n_{0}  
					\end{bmatrix}					
			\eea
where $\bbC=\prod_{r=1}^{3}\cC_{\Delta_r,l_r}$. 
In this case, happily we found exact one box tensor structure for each interaction vertex.

\subsection{The $(1,1,2)$ case}
\paragraph{}	
In the most general case involving three symmetric traceless fields with spins $l_{1,2}$ and $l_0$, as noted in \cite{3pt-Coupling0, 3pt-Coupling}, the corresponding three point ordinary Witten diagrams can only be expressed in terms of linear combination of box tensor basis \eqref{boxxx}.  
The same thing happens for the geodesic vertices in \eqref{Def:3ptVertex-geo} and the resultant three point geodesic Witten diagrams, they can only be expressed in terms of linear combination of box basis.
\paragraph{}
As an illustrative example, we consider the case where $(l_1, l_2, l_0) = (1, 1, 2)$. First from the corresponding CFT three point correlation function, 
we expect there are five box tensor structures arising, they are:
				\be \label{112Iboxes}
	[I_1]:=\begin{bmatrix}
					\Delta _1 &\Delta _2&  \Delta _0      \\[0.05em]
					1 & 1           & 2 \\[0.05em]
					0         &     0     & 0  
				\end{bmatrix},	
				[I_2]:=	\begin{bmatrix}
					\Delta _1 &\Delta _2&  \Delta _0      \\[0.05em]
					1 & 1           & 2 \\[0.05em]
					1         &     0     & 0  
				\end{bmatrix},
				[I_3]:=	\begin{bmatrix}
					\Delta _1 &\Delta _2&  \Delta _0      \\[0.05em]
					1 & 1           & 2 \\[0.05em]
					0         &     1     & 0  
				\end{bmatrix},
					 [I_4]:=	\begin{bmatrix}
					\Delta _1 &\Delta _2&  \Delta _0      \\[0.05em]
					1 & 1           & 2 \\[0.05em]
					1         &     1     & 0  
				\end{bmatrix},
				[I_5]:=	\begin{bmatrix}
					\Delta _1 &\Delta _2&  \Delta _0      \\[0.05em]
					1 & 1           & 2 \\[0.05em]
					0         &     0     & 1  
				\end{bmatrix}.
				\ee 
From the vertex parameterization \eqref{Def:3ptVertex-geo}, we now also have five independent interaction vertices. 
Let us denote the integral for the resultant three point geodesic Witten diagram for each vertex by $\left[J_{1, 1, 2}^{\rn_{1}, \rn_{2}, \rn_{0}}\right]$. 	
The order of $\{\rn_1, \rn_2, \rn_0\}$ we pick is 
\be\label{112Jboxes}
\left[J_1\right]:=\left[J_{1, 1, 2}^{0,0,0}\right], \left[J_2\right]:=\left[J_{1, 1, 2}^{1,0,0}\right], \left[J_3\right]:=\left[J_{1, 1, 2}^{0,1,0}\right], \left[J_4\right]:=\left[J_{1, 1, 2}^{1,1,0}\right], \left[J_5\right]:=\left[J_{1, 1, 2}^{0,0,1}\right].
\ee
The actual calculations producing them are complicated but somehow mechanical, 
however we can keep using the recursive relations of for the anti-symmetric tensor $\rC_{iAB}$ listed in Appendix \ref{rewritting-ts} to show that they can all be expressed in terms of box tensor structures given in \eqref{112Iboxes}.
\paragraph{} 
We can express the final results through the following matrix multiplication: $[J_a]=\mathbb{T}_{ab}[I_b]$,  $a, b =1, \dots, 5$
where the mixing matrices $\mathbb{T}_{ab}$ for simplified case $\Delta _2 = \Delta _1,\Delta_0 = \Delta$ is given by:			
\begin{equation} 
				\begin{array}{lcl}
				\mathbb{T}_{ab}=4\left(1+\Delta _1\right)\beta _{0,\Delta +2}\bbC\\\
				\begin{pmatrix}
				-\left(-4+\Delta ^2\right) \left(2+\Delta _1\right) &
				\frac{2 (2+\Delta ) \left(1+\Delta +\Delta _1\right)}{\Delta } & 2 (2+\Delta ) \left(2+\Delta _1\right) &\frac{2 (2+\Delta ) \left(1+\Delta +\Delta _1\right)}{\Delta } & 0\\[0.05em]
				-\Delta  & -1-\Delta  & -\frac{\Delta +\Delta ^2+2 \Delta _1}{\Delta +\Delta  \Delta _1} & -\frac{(1+\Delta ) \left(\Delta +\Delta _1\right)}{\Delta  \left(1+\Delta _1\right)} & 0
				\\[0.05em]
				-2+\Delta  & -2 & -1-\frac{2}{\Delta }+\Delta  & -\frac{1+\Delta }{\Delta } & 0 \\[0.05em]
				\frac{1}{1+\Delta _1} & \frac{1+\Delta }{\Delta +\Delta  \Delta _1} & \frac{1+\Delta }{\Delta +\Delta  \Delta _1} & \frac{1+\Delta }{\Delta +\Delta  \Delta _1} & 0\\[0.05em]
				0 & 0 & 0 & 0 & \Delta _1
				\end{pmatrix}
				\end{array}.
				\end{equation}
In particular, one can check that $\mathbb{T}$ is invertible such that:
				\be\text{Det}[\mathbb{T}_{ab}] \propto  \frac{ (-1+\Delta )^3 (2+\Delta )^2 \Delta _1^2 \left(1+\Delta _1\right){}^3 \left(2 (1+\Delta )^2+\left(2+2 \Delta +\Delta ^2\right) \Delta _1\right)}{\Delta ^3}\neq 0,
				\ee
This implies that we can equivalently express each three point function tensor structures listed in \eqref{112Iboxes} in terms of linear combination of three point GWDs for various vertices in \eqref{112Jboxes}.
This clearly illustrate that, the holographic dual of three point function for primary operators with spins, as expressed in the box tensor basis, generally requires more than one type of interaction vertices,   
and to find the ideal basis for two sets of quantities which give one to one correspondence, this essentially becomes a matrix diagonalization problem
\footnote{Here we should however mention here that in recent preprint \cite{Dyer1}, using the new CFT tensor basis constructed from linear combination of \eqref{Def:Hij} and \eqref{Def:Vijk}, and suitably constructed AdS space differential operators, the progress for direct identifications between CFT tensor structures and AdS interaction vertices has been made.}.
Moreover, recalling that we further can connect the box tensor basis appearing in \eqref{112Iboxes} with their corresponding differential tensor basis \eqref{differBasis}:  	
\bea
&&\{D_1\} := \begin{Bmatrix}
					\Delta _1 &\Delta _2&  \Delta _0      \\[0.05em]
					1 & 1           & 2 \\[0.05em]
					0         &     0     & 0  
				\end{Bmatrix},	
\{D_2\}:= \begin{Bmatrix}
					\Delta _1 &\Delta _2&  \Delta _0      \\[0.05em]
					1 & 1           & 2 \\[0.05em]
					1         &     0     & 0  
				\end{Bmatrix},
\{D_3\}:=  \begin{Bmatrix}
					\Delta _1 &\Delta _2&  \Delta _0      \\[0.05em]
					1 & 1           & 2 \\[0.05em]
					0         &     1     & 0  
				\end{Bmatrix},\nn\\
&&\{D_4\}:= \begin{Bmatrix}
					\Delta _1 &\Delta _2&  \Delta _0      \\[0.05em]
					1 & 1           & 2 \\[0.05em]
					1         &     1     & 0  
				\end{Bmatrix},
\{D_5\}:=\begin{Bmatrix}
					\Delta _1 &\Delta _2&  \Delta _0      \\[0.05em]
					1 & 1           & 2 \\[0.05em]
					0         &     0     & 1  
				\end{Bmatrix}. \label{112Dboxes}
\eea
Again for $\Del_1 = \Del_2$ and $\Del_0= \Del$, their mixing matrix is given by:
\be
\mathbb{A}_{ab}=\left(
					\begin{array}{ccccc}
					1-\frac{1}{4} \Delta  (4+\Delta ) & -\frac{\Delta }{2} & -\frac{\Delta }{2} & -\frac{1}{2} & \frac{2-\Delta }{4} \\
					-\frac{1}{4} (-2+\Delta ) \Delta  & \frac{\Delta }{2} & 1-\frac{\Delta }{2} & \frac{1}{2} & -\frac{\Delta }{4} \\
					-\frac{1}{4} (-2+\Delta ) \Delta  & 1-\frac{\Delta }{2} & \frac{\Delta }{2} & \frac{1}{2} & -\frac{\Delta }{4} \\
					-\frac{1}{4} (-2+\Delta )^2 & \frac{1}{2} (-2+\Delta ) & \frac{1}{2} (-2+\Delta ) & -\frac{1}{2} & \frac{2-\Delta }{4} \\
					0 & 0 & 0 & 0 & 1
					\end{array}
					\right)
\ee
such that $\{D_a\} =\bbA_{ab} [I_b]$, one can show that $\mathbb{A}_{ab}^{-1}$ is again invertible and agrees with Example 3.3.3 in \cite{SpinningBlock0} for $l=2$.					 
It should now be clear that, through two successive matrix multiplications, we can directly relate the differential tensor basis, 
which are somewhat more natural for constructing the integral representation of spinning conformal partial waves as explained in the previous section, to the three point GWDs for different interaction vertices.
We can succinctly summarize it as:
\be\label{DJ-relation}
\{D_a\} = (\bbA \bbT^{-1})_{ab} [J_b],
\ee
again it would be very interesting to find the new combination of interaction vertices which diagonalizes the matrix $\bbA\bbT^{-1}$, 
such that we can have the simple one to one correspondence with the CFT differential tensor basis.

\subsection*{Comments on Gluing Procedure}
\paragraph{}
So far, we have considered three point geodesic diagrams with a certain interaction.
Here we assume generic three point GWDs with external spins $(l_1,l_2,J)$ and an arbitrary interaction.
To use the gluing identity \eqref{Scalar-CPW-integral}, the dimension $\Delta_{0}$ is taken as $h+ i\nu$.\,\footnote{
For the right side diagram, it is taken as $h-i\nu$\,.}
After the geodesic integration, the resultant three point GWD is written in terms of the box tensor structures,
and we can reproduce the same box tensor structure using a summation of the differential operators as in \eqref{differBasis}\,.
Therefore we can write the following relation;
\be\label{diff-and-square}
\cD_{\rm Left}^{(l_1,l_2,J)}I_{\text{GWD}}^{(0,0,J)}=(\text{coeff.})~ I_{\text{GWD}}^{(l_1,l_2,J)} \,,
\ee
where $I_{\text{GWD}}^{(0,0,J)}$ is the three point GWD with $(0,0,J)$ external spins which computed 
in Section \ref{Section:Review} and $\cD_{\rm Left}^{(l_1,l_2,J)}$ is a linear summation of operators $\cD_{\rm Left}^{n_{10},n_{20},n_{12}}$ defined in \eqref{Def:DLeft}
which produces the same tensor structures as $I_{\text{GWD}}^{(l_1,l_2,J)}$.
The coefficient in RHS, denoted as  (coeff.) comes from the action of $\cD_{\rm Left}$\,.
$\cD_{\rm Left}$ produces only the Pochhammer symbols involving $\nu$ which do not give any additional poles when performing the $\nu$ integration.
For $I_{\text{GWD}}^{(0,0,J)}$\,, we already know how these two geodesic diagrams
can be glued together in Section \ref{Section:Review}, c. f. \eqref{Scalar-GWD-Split}.
If $\cD_{\rm Left}$ and $\cD_{\rm Right}$ act on the both side of \eqref{Scalar-CPW-integral}, in the RHS, we obtain the same
differential basis as in \eqref{diff-and-square}\,.
On the other hands, the LHS becomes the corresponding spinning
conformal partial wave.
In this way, we can concern the gluing process for an arbitrary pair 
of three point GWDs.											
\paragraph{}
Having illustrated how the three point interaction vertices parameterized in \eqref{Def:3ptVertex-geo} can be expressed in terms of the linear combination of box tensor basis, 
we can summarize the general strategy for constructing four point spinning GWDs which are holographic dual to the spinning conformal partial wave 
listed in \eqref{Def:SpinningBasis} as follows:
\begin{enumerate} 
\item{First consider a pair of triplets of CFT primary operators with scaling dimensions and spins $(\Del_{1,2}, l_{1,2})$ and $(h+i\nu, J)$ and $(l_{3,4}, \Del_{3,4})$ and $(h-i\nu, J)$, compute all the resultant three point spinning GWDs for a given pair of vertices parameterized \eqref{Def:3ptVertex-geo}, and express them in terms of the linear combination box tensor basis, i. e. working out the $\bbT$-matrix.}
\item{For each box tensor basis appearing, we further rewrite them into corresponding differential tensor basis, i.e. working out the $\bbA$ matrix.}
\item{We can next fuse the resultant differential basis together to obtain the direct relation between the four point spinning GWDs constructed from this pair of three point vertices and the spinning conformal partial waves.}
\item{Finally, if we consider all possible pairs of interaction vertices for the operators involved, and repeat the steps 1,2,3, we can then invert the relation between the spinning GWDs and spinning conformal partial waves, and express the spinning conformal partial waves in terms of linear combination of spinning GWDs instead.}
\end{enumerate}

\section{Decomposition of Witten Diagrams via Split Representation}
\label{Section:Decomposition}
\paragraph{}
In this section, we discuss how to decompose both four point scalar and spinning Witten diagrams involving general spin-$J$ exchange into four point geodesic Witten diagrams for the single and double trace operators. 
{The original analysis of decomposition have been done in \cite{ScalarGWD} for $J=0, 1$ exchanges,
the analysis we perform here rely on the so-called ``split representation'' of the bulk to bulk propagator introduced in \cite{SpinningAdS}},
and this makes clear why we can naturally construct various four point geodesic Witten diagrams from the three point ones, 
and their connection with the integral representation of conformal block itself.
One can regard the cutting identity \eqref{bulk-boundary relation} which was used in the previous sections as the natural consequence of the split representation.
\paragraph{}
We should clarify here that the analysis in this section can be regarded as a recasting the conformal partial wave decompositions of the four point ordinary scalar Witten diagrams done in \cite{SpinningAdS, SpinningAdS2, Sleight-Thesis} directly in terms of geodesic Witten diagrams. 
To do so we precisely identify the three point GWD contributions in the resultant split representation, while the remaining factors determine the spectrum of exchanged operators, 
the computational details can be found in Appendix \ref{App:Decomposition}.
We will see this somewhat easier approach, which is different from the one used in \cite{ScalarGWD},  
directly leads to the decomposition of ordinary Witten diagrams into GWDs for arbitrary spin $J$ and it is easier to generalize to the Witten diagrams for external operators with spins\footnote{We are grateful to Charlotte Sleight, whose comments encouraged us to explain our intentions better.}.

\paragraph{}
It was shown in \cite{SpinningAdS} the bulk to bulk propagator \eqref{Def:spinJPibb}  in so-called traceless gauge\footnote{One should note that bulk-bulk propagator for spin-$J$ tensor field can also be expressed in other gauge choice \cite{SpinningAdS2}.}, can be expressed as:
\bea\label{Def:Split-Rep}
\Pi_{\Del, J}(X, \tX; W, \tW)
&=& \sum_{l=0}^J \int^{\infty}_{-\infty} d\nu {a}_{l}(\nu) (W\cdot\nabla)^{J-l} ( \tilde{W}\cdot \tilde{\nabla})^{J-l}\Omega_{\nu, l}(X,\tilde{X}; W, \tilde{W})\,.\\
\Omega_{\nu, l}(X,\tilde{X}; W, \tilde{W})
&=& \frac{\nu^2}{\pi l!(h-1)_{l}} \int_{\partial} dP_0\,  \Pi_{h+i\nu, l}(X, P_0; W, D_{Z_0})\Pi_{h-iv, l}(\tX, P_0; \tilde{W}, Z_0)\,.\no\\
\label{Def:Split-Harmonic}
\eea
Here the embedding space covariant derivative $\nabla_A$ (or $\tilde{\nabla}_{\tA}$) is defined in \eqref{Def:CovDerivative}, and it satisfies properties $X^A\nabla_A=0$ and $\nabla_{A}G_{BC} = 0$. 
The function $\Omega_{\nu, l}(X, \tX; W, \tW)$ is the spin-$l$ harmonic in ${\rm AdS}_{d+1}$ space, 
$P_0$ and $Z_0$ denote the coordinate of the boundary point to be integrated over and its auxiliary polarization vector.
The key feature of the representation here is that we have expressed the AdS-harmonic functions in terms of the products of the bulk to boundary propagator $\Pi_{h\pm i\nu, l}(X, P_0; W, Z_0)$, 
hence the name ``split representation''.
Here the meromorphic functions $a_{l}(\nu)$, $l=0, 1, \dots, J$ have been obtained in  \cite{SpinningAdS} by comparing with the spectral functions in the conformal partial wave expansion of the corresponding CFT four point correlation function:
\bea\label{Def:coeff-a}
a_J(\nu) &=& \frac{1}{\nu^2+(\Del-h)^2}\,,\\
a_{l}(\nu)&=&\sum^{J-l}_{q=1}\frac{(l+q)!}{l! q!}\frac{(-1)^{q+1}}{2^{q-1}(q-1)!(h+l)_{q-1}}\frac{a_{l+q}(i(h-1+l))}{\nu^2+(h+l+q-1)^2}.
\eea
It is interesting to note that only $a_{J}(\nu)$ contains simple poles whose locations explicitly depend  on scale dimension $\Del$,
while $a_{l}(\nu)$ for $l < J$ are determined recursively by demanding the cancelation of  the residues for spurious poles in CFT spectral functions.
\paragraph{}
\begin{figure}
\centering
\includegraphics[width=0.4\linewidth]{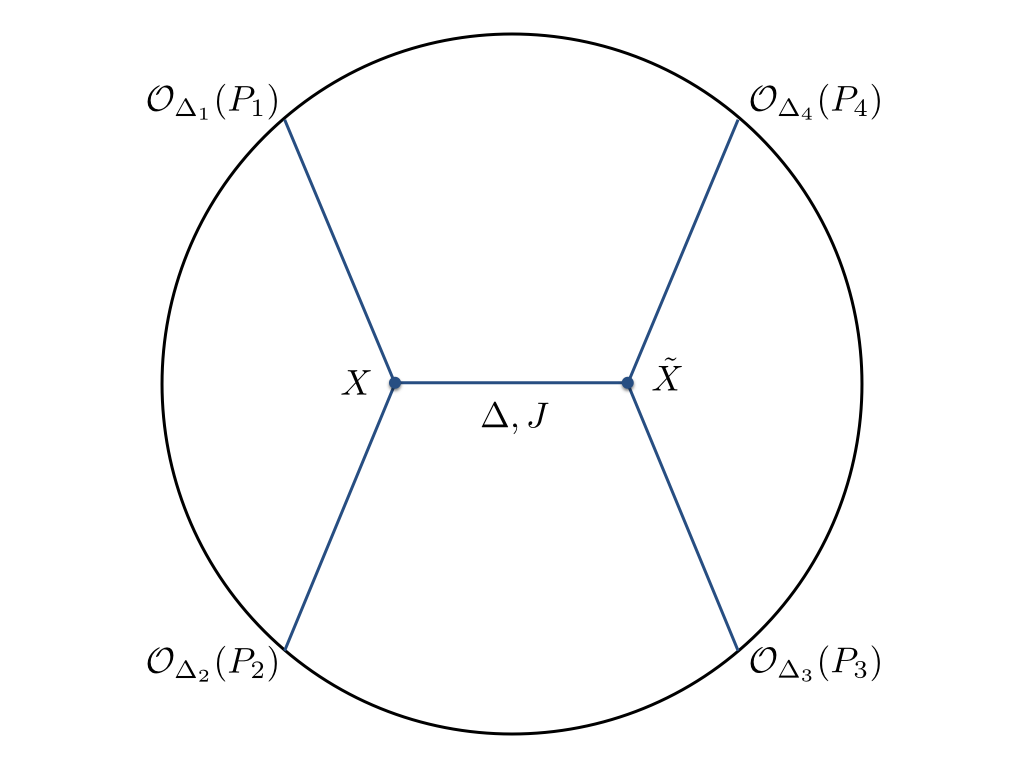}\qquad
\includegraphics[width=0.4\linewidth]{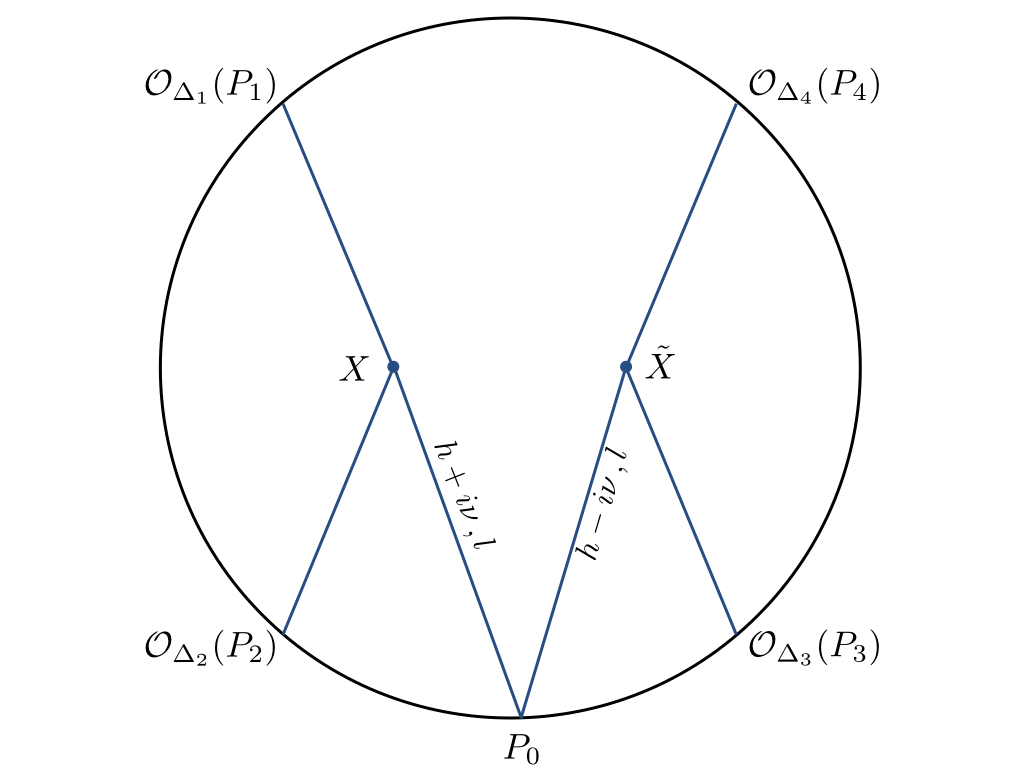}
\caption{Left: Normal exchange Witten diagram with four scalar external fields and a spin-$J$ internal field,
Right: A split diagram \label{Fig:4pt-Witten}}
\end{figure}
In the following, we will demonstrate how spin-$J$ exchange diagrams with scalar external fields are decomposed into
conformal partial waves/geodesic Witten diagrams.
The spin-$J$ operator exchange Witten diagram with 4 external scalar fields is given (See the left diagram in Figure \ref{Fig:4pt-Witten}):
\begin{eqnarray}\label{4-pt-NWD}
W^{\text{4-pt.}}_{(0,0),J,(0,0)}&\equiv&\frac{1}{\left(J! \left(\frac{d-1}{2}\right)_J\right)^2} \int \! dX d\tilde{X} \frac{1}{(-2P_1\cdot X)^{\Delta_1}}(K\cdot\nabla_X)^J \frac{1}{(-2P_2\cdot X)^{\Delta_2}}  \no\\
&&\times \frac{1}{(-2P_3\cdot \tilde{X})^{\Delta_3}}(\tilde{K}\cdot \nabla_{\tilde{X}})^J\frac{1}{(-2P_4\cdot \tilde{X})^{\Delta_4}}\Pi_{\Delta,J} (X,\tilde{X};W, \tilde{W})\,.
\end{eqnarray}
Here we dropped unimportant normalization factors $\cC_{\Delta_i,0}$\,.
This diagram can be decomposed into product of three point Witten diagrams 
by using the split representation \eqref{Def:Split-Rep} for the bulk to bulk propagator $\Pi_{\Del, J}(X, \tX; W, \tW)$ (See the
right digram in Figure \ref{Fig:4pt-Witten}):
\begin{eqnarray}
&&W^{\text{4-pt.}}_{(0,0),J,(0,0)}=\sum^J_{l=0}\int_\partial \!dP_0 \int^\infty_{-\infty}\! d\nu~ \frac{\nu^2}{\pi}\frac{1}{l!\, (h-1)!}a_l(\nu)\\
&&\times\left[\frac{\mathcal{C}_{h+i\nu, l}}{J! \left(\frac{d-1}{2}\right)_J}\int\! dX \frac{1}{(-2P_1\cdot X)^{\Delta_1}}(K\cdot\nabla_X)^J\frac{1}{(-2P_2\cdot X)^{\Delta_2}} 
(W\cdot \nabla_X)^{J-l}\frac{\bigl(V_0(X,D_{Z_0})\cdot W\bigr)^l}{(-2P_0\cdot X)^{h+i\nu+l}}\right]\no\\
&&\times\left[\frac{\mathcal{C}_{h-i\nu,l}}{J! \left(\frac{d-1}{2}\right)_J}\int\! d\tilde{X}\frac{1}{(-2P_3\cdot \tilde{X})^{\Delta_3}}(\tilde{K}\cdot \nabla_{\tilde{X}})^J\frac{1}{(-2P_4\cdot \tilde{X})^{\Delta_4}}(\tilde{W}\cdot \nabla_{\tilde{X}})^{J-l} \frac{\bigl(V_0(\tilde{X},Z_0)\cdot \tilde{W}\bigr)^l}{(-2P_0\cdot \tilde{X})^{h-i\nu+l}} \right]\,,\no
\end{eqnarray}
where we have defined the vector:
\be\label{Def:V0}
V_0^A (X, Z) \equiv V_0^A (X) = 2(X\cdot \rC_0)^A = 2((X\cdot Z_0) P_0^A- (X\cdot P_0) Z^A).
\ee
We will concentrate on three point Witten diagrams in the square parentheses:
\begin{eqnarray}\label{3-pt-WD-I}
&&I_{(J,l)}^{\Delta_1,\Delta_2,h+i\nu}\equiv\\&& \frac{1}{J!\, \left(\frac{d-1}{2}\right)_J}\int_{\text{AdS}}\! dX~ \frac{1}{(-2P_1\cdot X)^{\Delta_1}}(K\cdot\nabla)^J \frac{1}{(-2P_2\cdot X)^{\Delta_2}}(W\cdot \nabla)^{J-l}  \frac{(W\cdot V_0(X))^l}{(-2P_0\cdot X)^{h+i\nu+l}} \no\,.
\end{eqnarray}
We can simplify the integrands involved as:
\begin{eqnarray}
 &&\frac{1}{J!\, \left(\frac{d-1}{2}\right)_J}(K\cdot\nabla)^J \frac{1}{(-2P_2\cdot X)^{\Delta_2}}(W\cdot \nabla)^{J-l}  \frac{(W\cdot V_0(X))^l}{(-2P_0\cdot X)^{h+i\nu+l}}\\
  &&=\sum^{J-l}_{p=0}{}_{J-l}C_p (\Delta_2)_J (h+i\nu+l)_{J-l} (-2 P_{20})^{J-l-p} \frac{\left(-2 P_2\cdot V_0 (X)\right)^{l}}{(-2P_2\cdot X)^{\Delta_2+J-p}(-2P_0\cdot X )^{h+i\nu +J-p}}\no\,,
\end{eqnarray}
where ${}_p C_{q} = \frac{p!}{(p-q)! q!}$ is the combinatorial factor in binomial expansion.
We can explicitly evaluate the integral \eqref{3-pt-WD-I} for these three point Witten diagrams as\footnote{
The same diagram is calculated in \cite{SpinningAdS} by using the series expansion of Gegenbauer polynomials.}
:
\begin{eqnarray}\label{3-pt-WD-I-2}
I_{(J,l)}^{\Delta_1,\Delta_2,h+i\nu}&=& \sum^{J-l}_{p=0}{}_{J-l}C_p\, \frac{{(-2)^l}(\Delta_2)_J (h+i\nu+l)_{J-l}}{(h+i\nu+J-l-p)_l} (-2 P_{20})^{J-l-p} \\
&&\qquad \times (\rD_{02})^l \int \! dX \frac{1}{(-2P_1\cdot X)^{\Delta_1}}\frac{1}{(-2P_2\cdot X)^{\Delta_2 +J-p}}
\frac{1}{(-2P_0\cdot X)^{h+i\nu +J-l-p}}\no\\
&=& \sum^{J-l}_{p=0}{}_{J-l}C_p\, \frac{{(-2)^l}(\Delta_2)_J (h+i\nu+l)_{J-l}}{(h+i\nu+J-l-p)_l} (-2 P_{20})^{J-l-p} 
 \mathcal{N}^{\Delta_1, \Delta_2 +J-p,h+i\nu+J-l-p}\no\\
&& \qquad \times (\rD_{02})^l \hat{\cA}^{\Delta_1, \Delta_2 +J-p,h+i\nu+J-l-p}\no\\
&=&
\sum^{J-l}_{p=0}{}_{J-l}C_p\, \frac{(-2)^{{J-p}} (-1)^{l}\pi^h (\Delta_2)_J}{\Gamma(\Delta_1) \Gamma(\Delta_2+J-p)} (h+i\nu+J-p)_{p}~ \Gamma\left( \frac{\Delta_1+\Delta_2-h\pm i\nu +l}{2}\right)\no\\
&&~\times \left(\frac{\Delta_1+\Delta_2-h+i\nu+l}{2}\right)_{J-l-q}
\left(\frac{-\Delta_{12}+h+i\nu +l}{2}\right)_{J-l-q}\no\\
&&~\times \beta_{\Delta_{12},h+i\nu+l}
\begin{bmatrix}
			\Delta _1 &\Delta _2&  \frac{d}{2}+i\nu       \\[0.05em]
			0 & 0           & l \\[0.05em]
			0         &     0     & 0  
			\end{bmatrix}\,.\no
\end{eqnarray}
{Here the factor $\cN^{\Del_1,\Del_2, \Del_3}$ and $\hat{\cA}$ in the second line are defined in \eqref{Def:N123}} and derivative operator $\rD_{02}$ is defined as 
\begin{eqnarray}
\rD_{02}\equiv (Z_0\cdot P_2)\left(Z_0\cdot \frac{\partial}{\partial Z_0}-P_0\cdot \frac{\partial}{\partial P_0}\right)+(P_0\cdot P_2) \left(Z_0\cdot \frac{\partial}{\partial P_0}\right)\,,
\end{eqnarray}
and satisfies useful identitiy:
\begin{eqnarray}
(\rD_{02})^l\frac{1}{(-2P_0\cdot X)^a}&=&(a)_l\frac{({P_2\cdot V_0 (X))^l}}{(-2 P_0\cdot X)^{a+l}}\,,\no
\end{eqnarray}
In the second equality in \eqref{3-pt-WD-I-2}, we used the result in Appendix \ref{App:normal-Witten},
and in the last equality in \eqref{3-pt-WD-I-2}, note that we used the notation $\Gamma(a\pm b) = \Gamma(a+b)\Gamma(a-b)$ for simplicity.
Moreover in the last line we have also isolated the piece which can identified with the integrated results from three point GWD, c. f. \eqref{3pt-ScalarGWD-J} or more generally \eqref{3pt-ScalarGWD-l}.
In Appendix \ref{App:Decomposition}, we make this identification more explicit through direct computations.
\paragraph{}
Now moving to perform the decomposition analysis, we need to consider gluing the product of three point Witten diagrams we just evaluated together by integrating over the spectral parameter $\nu$,
the singularity structure of the $\nu$-dependent function multiplying the three point GWD piece crucially determines possible spectrum of the four point GWDs or equivalently scalar conformal blocks can appear.
The original four point Witten diagram \eqref{4-pt-NWD} can be expressed as:
\begin{eqnarray}\label{4-pt-NWD-2}
&&W^{\text{4-pt.}}_{(0,0),J,(0,0)}=\sum^J_{l=0}\int_\partial \!dP_0 \int^\infty_{-\infty}\! d\nu~ \frac{a_l(\nu)}{\pi\, l!\, (h-1)!}
\sum^{J-l}_{p=0}\sum^{J-l}_{p'=0}\Lambda^{J,l,p,p'}_{\Delta_{1,2,3,4}}(h+i\nu+J-p)_{p}
(h-i\nu+J-p')_{p'}\no\\
&&~~\times\mathcal{R}^{J,l,p}_{\Delta_1,\Delta_2,h+i\nu}\mathcal{R}^{J,l,p'}_{\Delta_3,\Delta_4,h-i\nu}
\Gamma\left(\frac{\Delta_1+\Delta_2-h\pm i\nu+l}{2}\right)\Gamma\left(\frac{\Delta_3+\Delta_4-h\pm i\nu+l}{2}\right)\no\\
&&~~\times \nu^2\,\mathcal{C}_{h+i\nu,l}\mathcal{C}_{h-i\nu,l}
\beta_{\Delta_{12},h+i\nu+l}\beta_{\Delta_{34},h-i\nu+l}
\begin{bmatrix}
				\Delta _1 &\Delta _2&  h+i\nu    \\[0.05em]
				0 & 0           & l \\[0.05em]
				0         &     0     & 0  
			\end{bmatrix}
\cdot			
\begin{bmatrix}
				\Delta _3 &\Delta _4&  h-i\nu  \\[0.05em]
				0 & 0           & l \\[0.05em]
				0         &     0     & 0  
			\end{bmatrix}\,\no\\
&&=\sum^J_{l=0}\int_\partial \!dP_0 \int^\infty_{-\infty}\! d\nu~ \frac{a_l(\nu)}{\pi\, l!\, (h-1)!}
\sum^{J-l}_{p=0}\sum^{J-l}_{p'=0}\Lambda^{J,l,p,p'}_{\Delta_{1,2,3,4}}(h\pm i\nu+l)_{J-l}\no\\
&&~~\times\frac{\mathcal{R}^{J,l,p}_{\Delta_1,\Delta_2,h+i\nu}\mathcal{R}^{J,l,p'}_{\Delta_3,\Delta_4,h-i\nu}}{ (h+ i\nu+l)_{J-l-p}(h- i\nu+l)_{J-l-p'}   }
\Gamma\left(\frac{\Delta_1+\Delta_2-h\pm i\nu+l}{2}\right)\Gamma\left(\frac{\Delta_3+\Delta_4-h\pm i\nu+l}{2}\right)\no\\
&&~~\times{\mathcal{B}_{h,l}(\nu)\mathcal{K}_{\Delta_{12},\Delta_{34},l}(h+i\nu,h-i\nu)}
\begin{bmatrix}
				\Delta _1 &\Delta _2&  h+i\nu    \\[0.05em]
				0 & 0           & l \\[0.05em]
				0         &     0     & 0  
			\end{bmatrix}
\cdot			
\begin{bmatrix}
				\Delta _3 &\Delta _4&  h-i\nu  \\[0.05em]
				0 & 0           & l \\[0.05em]
				0         &     0     & 0  
			\end{bmatrix}\,.
\end{eqnarray}
Let us unpack the various contributions appearing above.
Here $\Lambda^{J,l,p,p'}_{\Delta_{1,2,3,4}}$ is a factor which does not depend on $\nu$:
\begin{eqnarray}
\Lambda^{J,l,p,p'}_{\Delta_{1,2,3,4}}\equiv\frac{{}_{J-l}C_p{}_{J-l}C_{p'}}{l!\, (h-1)!}\frac{(-2)^{{2J-p-p'}}\pi^{d-1} (\Delta_2)_J(\Delta_4)_J}{\Gamma(\Delta_1)\Gamma(\Delta_2-J-p)\Gamma(\Delta_3)\Gamma(\Delta_4-J-p')}\,,
\end{eqnarray}\label{Def:coeff-R}
and $\mathcal{R}^{J,l,p}_{\Delta_1,\Delta_2,h+i\nu}$ is also a regular function of $\nu$:
\begin{eqnarray}
\mathcal{R}^{J,l,p}_{\Delta_1,\Delta_2,h+i\nu}\equiv\left(\frac{\Delta_1+\Delta_2-h+i\nu+l}{2}\right)_{J-l-p}
\left(\frac{-\Delta_{1}+\Delta_{2}+h+i\nu +l}{2}\right)_{J-l-p}\,.
\end{eqnarray}
$\mathcal{R}^{J,l,p'}_{\Delta_3,\Delta_4,h-i\nu}$ is also defined in the similar way.
Finally we notice that the last line of \eqref{4-pt-NWD-2} is almost identical to the integrand appearing in the integral representation of four point scalar GWD \eqref{Scalar-CPW-integral} for spin $l$, $l=0, 1, \dots, J$, except now we need to carefully examine the pole structures multiplying it. 
We will again start with the relation \eqref{Key-Relation1} with $J \to l$, 
but now multiply both sides with the following factor and integrate over $\nu$:
\begin{eqnarray}
&&a_l(\nu)(h\pm i\nu+l)_{J-l}\mathcal{R}^{J,l,p}_{\Delta_1,\Delta_2,h+i\nu}\mathcal{R}^{J,l,p'}_{\Delta_3,\Delta_4,h-i\nu}
\frac{\mathcal{B}_{h,l}(\nu)}{(h+ i\nu+l)_{J-l-p}(h- i\nu+l)_{J-l-p'}}\no\\
&&\qquad \qquad \times\Gamma\left(\frac{\Delta_1+\Delta_2-h\pm i\nu+l}{2}\right)\Gamma\left(\frac{\Delta_3+\Delta_4-h\pm i\nu+l}{2}\right)\,.
\end{eqnarray}
The LHS is what we have in the summation of \eqref{4-pt-NWD-2} except for some overall constant factors, the RHS becomes
\begin{eqnarray}\label{GWDandCB}
&&\int^{\infty}_{-\infty} \! d\nu~ a_l(\nu)~(h\pm i\nu+l)_{J-l}
\frac{\mathcal{R}^{J,l,p}_{\Delta_1,\Delta_2,h+i\nu}\mathcal{R}^{J,l,p'}_{\Delta_3,\Delta_4,h-i\nu}}
{(h+ i\nu+l)_{J-l-p}(h- i\nu+l)_{J-l-p'}}\\
&&\quad \times\frac{1}{2^l \,c_l}\Gamma\left(\frac{\Delta_1+\Delta_2-h\pm i\nu+l}{2}\right)\Gamma\left(\frac{\Delta_3+\Delta_4-h\pm i\nu+l}{2}\right)\no\\
&&\quad \times\left[\frac{\mathcal{K}_{\Delta_{12},\Delta_{34},l}(h+i\nu ,h+i\nu)}{(h+i\nu-1)_l\Gamma (i\nu) \Gamma (h+i\nu +l)}W_{\mathcal{O}_{h+i\nu,l}}+\frac{\mathcal{K}_{\Delta_{12},\Delta_{34},l}(h-i\nu ,h-i\nu)}{(h-i\nu-1)_l\Gamma (-i\nu) \Gamma (h-i\nu +l)}W_{\tilde{\mathcal{O}}_{h-i\nu,l}}\right]
\,.\no
\end{eqnarray}
In the following, we will focus on the integral above and perform the integration over $\nu$ separately for each $l$, since they spectral function $a_l(\nu)$ \eqref{Def:coeff-a} differs.

\subsubsection*{The Highest Spin $l=J$ Contribution}
\paragraph{}
Here we consider the contribution from the highest spin exchange $l=J$ case in \eqref{GWDandCB}\,.
Note that in this case, $p$ and $p'$ can be taken as only $p=p'=0$\,.
Then the integration in \eqref{GWDandCB} becomes
\begin{eqnarray}\label{highest-spin-int}
&&\int^{\infty}_{-\infty} \! d\nu~ a_J(\nu)~\frac{1}{2^J \,c_J}\Gamma\left(\frac{\Delta_1+\Delta_2-h\pm i\nu+J}{2}\right)\Gamma\left(\frac{\Delta_3+\Delta_4-h\pm i\nu+J}{2}\right)\\
&&\quad \times\left[\frac{\mathcal{K}_{\Delta_{12},\Delta_{34},J}(h+i\nu ,h+i\nu)}{(h+i\nu-1)_J\Gamma (i\nu) \Gamma (h+i\nu +J)}W_{\mathcal{O}_{h+i\nu,J}}+\frac{\mathcal{K}_{\Delta_{12},\Delta_{34},J}(h-i\nu ,h-i\nu)}{(h-i\nu-1)_J\Gamma (-i\nu) \Gamma (h-i\nu +J)}W_{\tilde{\mathcal{O}}_{h-i\nu,J}}\right]
\,.\no
\end{eqnarray}
As in the calculation in Section \ref{Section:Review}, the conformal partial wave $W_{\mathcal{O}_{h+i\nu}}$ converges in the lower half plane in the $\nu$ integration,
for the first term the integration contour should be taken in the lower half plane. In the second term, the contour is taken in the upper half plane for the same reason.
In the first term, $a_J(\nu)$ and gamma functions in the first line have the following poles in the lower half plane:
\begin{eqnarray}
\nu&=&-i(\Delta-h)\,,\label{pole-from-a}\\
\nu&=&-i(\Delta^{(12)}_{m,J}-h)\,,\qquad 
\nu=-i(\Delta^{(34)}_{m,J}-h)\,, \qquad m=0,1,2,\dots \label{pole-from-g}
\end{eqnarray}
where the pole in \eqref{pole-from-a} comes from the coefficient $a_J(\nu)$ and the poles in \eqref{pole-from-g} 
come from gamma functions.
Here $\Delta^{(12)}_{m,J}$ is defined as
\begin{eqnarray}\label{Def:dim-double}
\Delta^{(12)}_{m,J}\equiv \Delta_1+\Delta_2+J+2m\,,
\end{eqnarray}
and $\Delta^{(34)}_{m,J}$ is also defined in the similar way.
These poles contribute the integration in the first term.
\eqref{pole-from-a} corresponds to the contribution from the single trace operator exchange 
and it is precisely the origin of the integral representation of scalar GWD \eqref{Scalar-GWD-Split}. 
While the remaining poles \eqref{pole-from-g} corresponds to the double trace operator exchange.
Similarly, in the second term, the following poles contribute;
\begin{eqnarray}
\nu&=&+i(\Delta-h)\,,\\
\nu&=&+i(\Delta^{(12)}_{m,J}-h)\,,\qquad 
\nu=+i(\Delta^{(34)}_{m,J}-h)\,, \qquad m=0,1,2,\dots
\end{eqnarray}
Then combining the contribution from the first and second term and multiplying some constant factors to \eqref{highest-spin-int},
we can obtain the following decomposition,
\begin{eqnarray}\label{decom:l=J}
\left.W^{\text{4-pt.}}_{(0,0),J,(0,0)}\right|_{l=J}= \tilde{\alpha}_{\Delta,J} W_{\mathcal{O}_{\Delta,J}}+\sum_{m=0}^\infty
\tilde{\alpha}_{\Delta^{(12)}_{m,J},J} W_{\mathcal{O}_{\Delta^{(12)}_{m,J},J}}+\sum_{m=0}^\infty
\tilde{\alpha}_{\Delta^{(34)}_{m,J},J} W_{\mathcal{O}_{\Delta^{(34)}_{m,J},J}}\,,
\end{eqnarray}
where each coefficient is determined by the residue of the corresponding poles.
Each conformal partial wave $W_{\cO_{\Del, J}}$, $W_{\cO_{\Del^{(12)}_{m, J},J}}$ and $W_{\cO_{\Del^{(34)}_{n, J},J}}$ in the RHS is proportional to the corresponding GWDs as in Section \ref{Section:Review}.
This expansion leads to the following decomposition into GWD $\cW$,
\begin{eqnarray}\label{decom:l=J2}
\left.W^{\text{4-pt.}}_{(0,0),J,(0,0)}\right|_{l=J}= \alpha_{\Delta,J} \cW_{\mathcal{O}_{\Delta,J}}+\sum_{m=0}^\infty
\alpha_{\Delta^{(12)}_{m,J},J} \cW_{\mathcal{O}_{\Delta^{(12)}_{m,J},J}}+\sum_{m=0}^\infty
\alpha_{\Delta^{(34)}_{m,J},J} \cW_{\mathcal{O}_{\Delta^{(34)}_{m,J},J}}\,.
\end{eqnarray}

\subsubsection*{The Spin $l < J$ Contributions}
\paragraph{}
Next we consider the $0 \le l<J$ cases in \eqref{GWDandCB}.
Let us first consider the contributions from the double trace operators, as encoded within the $\Gamma$-functions in the second line of \eqref{GWDandCB}.
In the first term, after the $\nu$-integration, non-vanishing residues arise from the poles at
\begin{eqnarray}\label{DoublePoles12}
\nu&=&-i(\Delta^{(12)}_{m,l}-h)\,,\qquad 
\nu=-i(\Delta^{(34)}_{m,l}-h)\,, \qquad m=0,1,2,\dots
\end{eqnarray}
and in the second term, the the poles at 
\begin{eqnarray}\label{DoublePoles34}
\nu&=&+i(\Delta^{(12)}_{m,l}-h)\,,\qquad 
\nu=+i(\Delta^{(34)}_{m,l}-h)\,, \qquad m=0,1,2,\dots
\end{eqnarray}
give similar residues \footnote{Note that in fact some of the poles are canceled by zeros of the Pochhammer symbols in \eqref{Def:coeff-R}\,.}.
After integrating over $\nu$ and multiplying some constants, we obtain the following decomposition for the lower spin $l<J$ case; 
\begin{eqnarray}\label{decom:l<J}
\left.W^{\text{4-pt.}}_{(0,0),J,(0,0)}\right|_{l<J}= \sum_{m=0}^\infty
\alpha_{\Delta^{(12)}_{m,l},l} \cW_{\mathcal{O}_{\Delta^{(12)}_{m,l},l}}+\sum_{m=0}^\infty
\alpha_{\Delta^{(34)}_{m,l},l} \cW_{\mathcal{O}_{\Delta^{(34)}_{m,l},l}}\,.
\end{eqnarray}
Here the coefficients come from the residues corresponding to each double trace poles.
\paragraph{}
Together with the result of the highest spin case \eqref{decom:l=J} and the lower spin case \eqref{decom:l<J},
the normal four point exchange diagram with a spin-$J$ internal field can be decomposed as;
\begin{eqnarray}\label{decom-scalar}
W^{\text{4-pt.}}_{(0,0),J,(0,0)}=\alpha_{\Delta,J} \cW_{\mathcal{O}_{\Delta,J}}
+\sum_{l=0}^{J}\left( \sum_{m=0}^\infty \alpha_{\Delta^{(12)}_{m,l},l} \cW_{\mathcal{O}_{\Delta^{(12)}_{m,l},l}}+\sum_{m=0}^\infty
\alpha_{\Delta^{(34)}_{m,l},l} \cW_{\mathcal{O}_{\Delta^{(34)}_{m,l},l}}\right)\,.
\end{eqnarray}
\paragraph{}
However to complete our decomposition analysis here, we notice an essential difference for $l<J$ cases is that
there can be so-called ``spurious pole contributions'' arises  \cite{SpinningAdS, ConformalRT},  from $a_{l<J}(\nu)$ as defined in \eqref{Def:coeff-a}.
For fixed $l$, they are located at:
\begin{eqnarray}
\nu=\pm i (h+l+q-1)\,.\qquad q=1,\dots, J-l
\end{eqnarray}
along the imaginary axis in complex $\nu$-plane, such that when we close contour in either lower or upper half plane in \eqref{GWDandCB},
they give four point GWDs/conformal partial waves associated with integer scaling dimensions which do not depend on  $\Delta_i$ or $\Delta$.
To illustrate these contributions are unphysical,
 consider the relations \eqref{Def:Split-Rep} and \eqref{Def:AdSHarm},
and the following consistency relation is obtained:
\begin{eqnarray}\label{ConsistentRel}
&&\Pi_{\Delta,J}(X,\tilde{X};W,\tilde{W})=\\
&&\qquad\sum_{l=0}^{J}\int_{-\infty}^{\infty}d\nu ~a_l(\nu) (W\cdot \nabla)^{J-l}
(\tilde{W}\cdot \tilde{\nabla})^{J-l}\left(\Pi_{h+i\nu,l}(X,\tilde{X};W,\tilde{W})-\Pi_{h-i\nu,l}(X,\tilde{X};W,\tilde{W})\right)\,.\no
\end{eqnarray}
After the $\nu$ integration, from the highest spin term $l=J$,
we can obtain the original bulk to bulk propagator,
therefore the remaining $l=0,1, \dots, J-1$ summation which only pick up residues from spurious poles must sum to zero.
From this point of view, the spurious pole contributions give no physical contributions.
However, when we substitute the split representation \eqref{Def:Split-Rep}
into the four point Witten diagram, we have performed the $X$ and $\tilde{X}$ integration first (or equivalently $\lambda$ and $\lambda'$), 
before performing $\nu$-integration, there are additional poles such as the double trace operators poles listed in \eqref{DoublePoles12} and \eqref{DoublePoles34} plus regular $\nu$-dependences appearing. In other words, $X$ and $\tilde{X}$ integrations do not commute with $\nu$ integration as should be expected. 
However crucially for our integrand  \eqref{4-pt-NWD}, these additional poles do not coincide with the spurious poles or affect convergence of subsequent $\nu$-integration,
as far as the final residues arising from the spurious poles are concerned, the $\nu$ integration commutes with $X$ and $\tilde{X}$ integrations.
We can thus use the \eqref{ConsistentRel} to argue that the residues arising from the spurious poles in \eqref{GWDandCB}, when we sum over all the $l=0, \dots, J-1$ contributions, 
should total to zero. This slightly simplified argument is in accord with the recursive relations imposed on $a_l(\nu)$  \cite{SpinningAdS}, 
which in turns arise from the cancelation of the spurious residues in the dual Mellin amplitude \cite{ConformalRT}.
This completes our generalization of the decomposition for four point scalar Witten diagrams into four point scalar GWDs done in \cite{ScalarGWD} for $J=0, 1$ to arbitrary $J$.
\paragraph{}
To close this section, we would like to consider possible Mellin representation \cite{PenedonesMellinAmp} of scalar GWDs.
The Mellin representation of CPWs is already written in \cite{DO-2011,ConformalRT}, which can be identified with its integral representation obtained from two copies of three point functions \eqref{Scalar-CPW-integral} hence the three point GWDs \eqref{Scalar-GWD-Split}, explicitly we have the following relation:
\begin{eqnarray}\label{MellinCPW}
W_{\Delta.J}(u,v)&\propto& \int^\infty_{-\infty}\! d\nu~ \frac{\mathcal{B}_{h,J}(\nu)}{\nu^2+(\Delta-h)^2}
\int^{i\infty}_{-i\infty}\frac{dt ds}{(4\pi i)^2} u^{\frac{t}{2}}v^{\frac{-(s+t)}{2}}P_{\nu,J}(s,t) (h\pm i\nu-1)_J\nn\\
&&\times \Gamma\left(\frac{h\pm i\nu -J-t}{2}\right)\Gamma\left(\frac{t+s}{2}\right)\Gamma\left(\frac{t+s+\Delta_{12}-\Delta_{34}}{2}\right) \Gamma\left(\frac{-\Delta_{12}-s}{2}\right)\Gamma\left(\frac{\Delta_{34}-s}{2}\right)   \no\,,\\
&\propto&   \int^{\infty}_{-\infty} d\nu  \frac{{\mathcal B}_{d, J}(\nu)} {\nu^2+(\Del-h)^2  } { \cK_{\Del_{12}, \Del_{34};J}\left(h+i\nu, h-i\nu\right) }
\int dP_0
\begin{bmatrix}
				\Delta _1 &\Delta _2&  h+i\nu    \\[0.05em]
				0 & 0           & J \\[0.05em]
				0         &     0     & 0  
			\end{bmatrix}	
\cdot	
\begin{bmatrix}
				\Delta _3 &\Delta _4&  h-i\nu  \\[0.05em]
				0 & 0           & J \\[0.05em]
				0         &     0     & 0  
			\end{bmatrix},\nn\\
\end{eqnarray}
where $s, t$ are Mellin integration variables and $P_{\nu, J} (s,t)$ is the Mack polynomial which is defined in the Appendix B in \cite{ConformalRT}.
It is known that \cite{Factorization} that Mellin amplitudes exhibits factorization properties when considering the residues associated with the infinite sequence of simple poles located at:
\be
t = h+i\nu-J+2m, \quad m =0, 1, 2, \dots
\ee
and also their shadows with $+i\nu \to -i\nu$. Such that for each spin-$J$ exchange, 
we can express the residues in terms of the lower point Mellin amplitudes, joined together by certain function which in flat space limit can be identified with the propagator of spin-$J$ particle.
In the simplest non-trivial case, we have four point Mellin amplitudes factorized into two copies of three point Mellin amplitudes, joined together by the ``propagator''.
Given the conformal partial waves are building blocks of four point correlation functions, its Mellin representation given in first line of \eqref{MellinCPW} inherits such a factorization, 
and the resultant pieces should be closely related to the building blocks of its holographic counterpart, i.e. three point GWDs, it would be very interesting to clarify such a relation.

\subsection*{Comments on fields with spins}
\paragraph{}
Here we consider the simple extension of Witten diagrams with external spinning fields.
The basic idea is to use the derivative operators defined in \eqref{Def:D11}-\eqref{Def:D21}.
A simple example can be obtain by using derivative operators $\rD_{12}$.
If we consider $\rD_{12}$ operator acting on the integration \eqref{3-pt-WD-I}, the following three point diagram appears:
\begin{eqnarray}\label{3-pt-WD(l-0-J)}
&&(\rD_{12})^{l_1}I_{b,(J,l)}^{\Delta_1,\Delta_2,h+i\nu}\\
&&=(\rD_{12})^{l_1}\frac{1}{J!\, \left(\frac{d-1}{2}\right)_J}\int_{\text{AdS}}\! dX~ \frac{1}{(-2P_1\cdot X)^{\Delta_1}}(K\cdot\nabla)^J \frac{1}{(-2P_2\cdot X)^{\Delta_2}}(W\cdot \nabla)^{J-l}  \frac{(W\cdot V_0(X))^l}{(-2P_0\cdot X)^{h+i\nu+l}}\no\\
&&=\frac{1}{J!\, \left(\frac{d-1}{2}\right)_J\,l_1!\, \left(\frac{d-1}{2}\right)_{l_1}}\frac{(\Delta_1)_{l_1}}{(\Delta_2)_{l_1}}\int_{\text{AdS}}\! dX~ \frac{(\tilde{K}\cdot V_1(X))^{l_1}}{(-2P_1\cdot X)^{\Delta_1+l_1}}
\no\\
&&\qquad \times 
(\tilde{W}\cdot\nabla)^{l_1}(K\cdot\nabla)^J \frac{1}{(-2P_2\cdot X)^{\Delta_2-l_1}}(W\cdot \nabla)^{J-l}  \frac{(W\cdot V_0(X))^l}{(-2P_0\cdot X)^{h+i\nu+l}} \no\,.
\end{eqnarray}
This corresponds to $(l_1,l_2,l_0)=(l_1,0,J)$ case with an interaction like
\begin{eqnarray}\label{int-(l,0,J)}
T_{(l_1)}^{\mu_1,...,\mu_{l_1}}\nabla_{\mu_1,...,\mu_{l_1},\nu_1,...,\nu_J}\Phi~ T_{(J)}^{\nu_1,...,\nu_J}\,.
\end{eqnarray}
After calculating the integration in \eqref{3-pt-WD(l-0-J)} with $\rD_{12}$, 
the integral is proportional to the following differential basis:
\begin{eqnarray}\label{DiffBasisl1l}
\begin{Bmatrix}
				\Delta _1 &\Delta _2&  h+i\nu  \\[0.05em]
				l_1 & 0           & l \\[0.05em]
				0         &     0     & 0  
\end{Bmatrix}\,,
\end{eqnarray}
as the result of the integration \eqref{3-pt-WD(l-0-J)} should be same as the last line of 
\eqref{3-pt-WD-I-2} acting with $(\rD_{12})^{l_1}$\,.
The coefficient in front of this basis \eqref{DiffBasisl1l}
can have $\nu$-dependence through the action of the derivative $\rD_{12}$ acting on $\hat{\cA}^{\Delta_1,\Delta_2,h+i\nu}$,
however this only resulted in the Pochhammer symbol which does not give additional poles.
In this case, we can decompose the spinning Witten diagram with the interaction in \eqref{int-(l,0,J)} in a similar way and 
as in the scalar case, now the conformal partial waves in \eqref{decom-scalar} are changed as $(\rD_{12})^{l_1}W_{\cO}$ where $W_{\cO}$
can be the scalar conformal partial wave for single and double trace operators in the RHS of \eqref{decom-scalar}.
\paragraph{}
Even if we consider more general cases with arbitrary external spins and interaction, 
the 3-point integration can be done basically and the result should be written in terms of the box basis
for the same reason as in Appendix \ref{rewritting-ts}\footnote{
It is difficult to specify the box basis corresponding to 
an arbitrary interaction. It seem that we should consider on case-by-case basis.}.
Then after the same argument, we obtain the single trace and the double trace contribution from
the $\nu$ integration. The resulting decomposition is expanded in terms 
of spinning conformal partial waves like $\mathcal{D}_{\text{left}}\mathcal{D}_{\text{right}}W_{\cO}$\,.

\acknowledgments
This work was supported in part by National Science Council through the grant  104-2112-M-002 -004 -MY, Center for Theoretical Sciences at National Taiwan University. 
HYC would like to thank Kyoto University, Korea Institute for Advanced Study and Keio University for the opportunities to present part of this work and the hospitalities during its completion.
The work of HK is supported by the Japan Society for the Promotion of Science (JSPS) and by the Supporting Program for Interaction-based Initiative Team Studies (SPIRITS) from Kyoto University.

\appendix

\section{Appendix: Embedding Formalism}\label{Appendix:Embed}
\paragraph{}
In this appendix we review the essential details about the embedding space formalism for encoding the tensors in both euclidean $d+1$ dimensional Anti-de Sitter space and the 
$d$-dimensional euclidean space living on its boundary, this formalism is particularly convenient for studying AdS$_{d+1}$ / CFT$_{d}$ correspondence. 
It is useful to realize the common $SO(d, 1)$ isometry group of AdS$_{d+1}$ space and conformal group of its $d$-dimensional boundary as
the Lorentz group of a $d+2$ dimensional Minkowski space.
The essence of the embedding formalism is that we can realize the non-linear  isometry and conformal transformations of the lower dimensional spaces as the 
linear Lorentz transformation of the associated embedding space, this becomes beneficial when dealing with tensors.
\paragraph{}
In $d+2$ dimensional embedding space $\bM^{d+1, 1}$, the euclidean $AdS_{d+1}$ space is defined by the set of future directed unit vectors satisfying:
\begin{equation}\label{Def:AdS-embed}
{\rm AdS}_{d+1}~:~X\cdot X=\eta_{AB} X^A X^B= -1,  \quad \quad \eta_{AB} = {\rm diag}(-1, 1, \dots, 1), \quad \quad  X^{0}>1
\end{equation}
which can also be viewed as a $d+1$ dimensional hyperboloid, and we have set the radius of curvature to be 1.
We can parametrize the solutions to \eqref{Def:AdS-embed} explicitly in the light cone coordinates:
\bea
 (X^+, X^-, X^a) = \frac{1}{z}(1, y^2+z^2, y^{a}), \quad X\cdot X = -X^+X^- +\delta_{ab} X^a X^b, \quad a, b =0, \dots ,d-1\nn\\
\eea
in terms of the Poincare coordinates $x^\mu = (z, y^a)$ of AdS$_{d+1}$ space.
Towards the boundary AdS$_{d+1}$, the hyperboloid asymptotes to the light cone $X\cdot X = 0$, 
i. e. the conformal boundary $\bR^d$ is identified with the projective cone of light rays in the embedding space.
They are given by the homogeneous coordinates subjected to the projective identification:
\begin{equation}\label{Def:Rd-embed}
\bR^d~:~P\cdot P =0, \quad \quad P^A \sim \lambda P^A, \quad \lam \neq 0.
\end{equation}
In terms of Poincare coordinates, the boundary points up to projective identification above are parameterized as:
\be
(P^+, P^-, P^a) = (1, y^2, y^a).
\ee
Next we consider embedding physical tensor fields in AdS$_{d+1}$ and $\bR^d$ into embedding space $\bM^{d+1, 1}$. 
Explicitly, given an arbitrary rank-r tensor field in AdS$_{d+1}$ or $\bR^{d}$, they are related to their embedding space counterparts through the pull-back operations:
\be
{\rm AdS}_{d+1}~:~ {\mathcal{T}}_{\mu_1\dots \mu_r}^{\rm (AdS)}(x) = \frac{\pa X^{A_1}}{\pa x^{\mu_1}}\dots \frac{\pa X^{A_r}}{\pa x^{\mu_r}} T_{A_1 \dots A_r} (X),\quad
{\bR}^d~:~ {\cF}_{a_1 \dots a_r}^{(\bR)}(y) = \frac{\pa P^{A_1}}{\pa y^{a_1}} \dots  \frac{\pa P^{A_r}}{\pa y^{a_r}}  F_{A_1\dots A_r}(P).\label{PB-AdS/Euc-tensor}
\ee
In particular, the AdS$_{d+1}$ and $\bR^{d}$ metrics are given by:
\be\label{PB-metrics}
{\rm AdS}_{d+1}~:~g_{\mu\nu}^{\rm (AdS)} = \frac{\pa X^A}{\pa x^\mu} \frac{\pa X^B}{\pa x^\mu} \eta_{AB}, \quad \quad \quad \bR^d~:~ \delta_{ab}^{(\bR)} = \frac{\pa P^A}{\pa y^a} \frac{\pa P^B}{\pa y^b} \eta_{AB}.
\ee
However the pull-back operations defined in \eqref{PB-AdS/Euc-tensor} are surjective but not injective, 
in other words given a physical tensor in AdS$_{d+1}$ or $\bR^d$, they do not have a unique representative in the embedding space $\bM^{d+1, 1}$, 
but rather the embedding introduces redundant unphysical degrees of freedom. 
We can see this from the orthogonal conditions:
\be\label{OrthoCond}
X_A\frac{\pa X^A}{\pa x^\mu}\vline_{X\cdot X = -1} = 0, \quad P_A \frac{\pa P^A}{\pa y^a}\vline_{P\cdot P = 0} = 0,
\ee
we can see that any tensor components proportional to $X_{(A_1} H_{A_2 \dots A_r)}(X)$ and $P_{(A_1} H'_{A_2 \dots A_r)}(P)$ contained respectively in $T_{A_1\dots A_r}(X)$ and 
$F_{A_1\dots A_r}(P)$ vanish under the pull-back operations in \eqref{PB-AdS/Euc-tensor}, hence unphysical.
Geometrically we can regard these extra components as being normal to the hypersurface \eqref{Def:AdS-embed} and \eqref{Def:Rd-embed} respectively.
We can thus eliminate these unphysical redundant degrees of freedom in the embedding space tensors by further imposing the {\it transverse} condition:
\be\label{TransCond}
X^{A_1} T_{A_1\dots A_r}(X)\mid_{X\cdot X = -1} = 0, \quad P^{A_1} F_{A_1 \dots A_r}(P)\mid_{P\cdot P =0} =0
\ee
such that $T_{A_1 \dots A_r}(X)$ and $F_{A_1 \dots A_r}(P)$ only contain the components which are tangent to AdS$_{d+1}$ and $\bR^d$ respectively. 
These are the embedding representatives of the AdS$_{d+1}$ and $\bR^d$ tensor fields.
\paragraph{}
Moreover in the main text, we would like to consider symmetric traceless AdS$_{d+1}$ and $\bR^d$ tensor fields. To construct their representatives in embedding space $\bM^{d+1, 1}$ 
they need to be symmetric traceless also transverse (STT) from the discussion above,
let us first introduce the following generating polynomials:
\bea
T(X, W) &=& W^{A_1} \dots W^{A_r} T_{A_1\dots A_r}(X), \quad\quad X\cdot W  = W\cdot W =0,\label{T-GenFun}\\  
F(P, Z) &=& Z^{A_1}\dots Z^{A_{r}}F_{A_1.\dots A_{r}}(P), \quad\quad P\cdot Z = Z\cdot Z = 0.\label{F-GenFun}   
\eea
Here we have introduced the auxiliary vectors $W^A$ and $Z^A$,  $X\cdot W =0$ and $W\cdot W=0$
imply $T_{A_1 \dots A_r}(X)$ is defined up to equivalence $\sim X_{(A_1} H_{A_2 \dots A_r)}(X)+\eta_{(A_1A_2} S_{A_3\dots A_r)}(X)$, 
the contraction with $W^A$s only picks up the symmetric, traceless and transverse components. 
Similarly the properties of the auxiliary vector $P^A$ ensures the contraction only picks up the transverse and traceless (plus symmetric) components of $F_{A_1\dots A_r}(P)$.
It is worth however noting that under the rescaling $F_{A_1\dots A_r}(\lam P) = \lam^{-\Del}F_{A_1\dots A_r} (P), ~\lam >0$, it is a homogenous polynomial of degree $-\Del$.
\paragraph{}
To recover embedding space STT tensors representing symmetric traceless AdS$_{d+1}$ and $\bR^d$ tensors directly from \eqref{T-GenFun} and \eqref{F-GenFun}, 
it is convenient to define the operators $K_A$ and $D_A$ which act on the symmetric products of $W^A$ and $Z^A$ respectively as:
\bea
&&\frac{1}{r!\left(\frac{d-1}{2}\right)_r} K_{A_1} \dots K_{A_r} W^{B_1} \dots W^{B_r} = G_{\{A_1}^{B_1}\dots G_{A_r\}}^{B_r} = G_{(A_1}^{B_1} \dots G_{A_{r})}^{B_r} - {\rm traces},\label{K-Action}\\
&&\frac{1}{r!\left(\frac{d-2}{2}\right)_r} D_{A_1} \dots D_{A_r} Z^{B_1} \dots Z^{B_r} = \Pi_{a_1 \dots a_r}{}^{b_1 \dots b_r}\frac{\pa P_{A_1}}{\pa y_{a_1}} \dots \frac{\pa P_{A_r}}{\pa y_{a_r}}  
\frac{\pa P^{B_1}}{\pa y^{b_1}} \dots \frac{\pa P^{B_r}}{\pa y^{a_r}}\label{D-Action}   
\eea
where $(\dots)$ in the above implies total symmetrization of indices and
\be
\Pi_{a_1 \dots a_r}{}^{b_1 \dots b_r} = \delta^{b_1}_{(a_1} \dots \delta^{b_r}_{a_r)} - {\rm traces}, \quad \frac{\pa P^A}{\pa y^b} = (0, 2x_b, \delta^a_b).
\ee
In other words we obtain the manifestly symmetric, traceless and transverse tensorial projectors, and the resultant embedding space tensors
\be\label{Def:STT-AdSTensor-embed}
T_{\{A_1\dots A_r\}}(X) = G^{B_1}_{\{A_1} \dots G^{B_r}_{A_r\}} T_{B_1\dots B_r}(X)
\ee
\be \label{Def:STT-RTensor-embed}
F_{\{A_1 \dots A_r\}}(P) = \Pi_{a_1 \dots a_r}{}^{b_1 \dots b_r}\frac{\pa P_{A_1}}{\pa y_{a_1}} \dots \frac{\pa P_{A_r}}{\pa y_{a_r}}  
\frac{\pa P^{B_1}}{\pa y^{b_1}} \dots \frac{\pa P^{B_r}}{\pa y^{a_r}} F_{B_1\dots B_r}(P)
\ee
are the desired STT representatives of AdS$_{d+1}$ and $\bR^d$ tensors in the embedding space $\bM^{d+1, 1}$. 
For completeness, explicit expression for the operators $K_A$ and $D_A$ can be given in terms of following differential operators:
\bea\label{Def:KAop}
K_A &=& \frac{d-1}{2}\left(\frac{\pa}{\pa W^A}+X_A\left(X\cdot \frac{\pa}{\pa W}\right)\right) +\left(W\cdot \frac{\pa}{\pa W}\right)\frac{\pa}{\pa W^A}\nn\\
&+&X_A\left(W\cdot \frac{\pa}{\pa W}\right) \left(X\cdot \frac{\pa}{\pa W}\right) -\frac{1}{2} \left(\frac{\pa^2}{\pa W\cdot \pa W}+ \left(X\cdot \frac{\pa}{\pa W}\right)\left(X\cdot \frac{\pa}{\pa W}\right)\right),
\eea
\be\label{Def:DAop}
D_A = \left(\frac{d-2}{2}+Z\cdot\frac{\pa}{\pa Z}\right)\frac{\pa}{\pa Z^A} -\frac{1}{2} Z_A \frac{\pa^2}{\pa Z\cdot \pa Z},
\ee
however we mostly will not use these somewhat lengthy expressions in the main text, only the formal operations \eqref{K-Action} and \eqref{D-Action} will be sufficient. 
When the contracted embedding space tensor in the generating polynomial is already traceless and transverse, the action of $K_A$ simplifies to
\begin{equation}
K_{A}=\left(\frac{d-1}{2}+W \cdot \frac{\partial}{\partial W }\right)\frac{\partial}{\partial W^{A} }
\end{equation}
Finally, we can consider the embedding space representative of AdS$_{d+1}$ covariant derivative, it acts on the embedding space tensor satisfying the transverse condition \eqref{TransCond}, 
and the resultant tensor should remain so after its action.
The following differential operator in $\bM^{d+1, 1}$ satisfies such requirement:
\begin{equation}\label{Def:CovDerivative}
\nabla_{A}=\frac{\partial}{\partial X ^{A}}+X _{A}\left( X \cdot \frac{\partial}{\partial X }\right)+W _{A} \left(X \cdot \frac{\partial}{\partial W }\right) = G_A{}^B\frac{\pa}{\pa X^B} + W _{A} \left(X \cdot \frac{\partial}{\partial W }\right) 
\end{equation}
we can clearly see that $X^A \nabla_A = 0$, and moreover if the contracted tensor in \eqref{T-GenFun} already satisfies the transverse condition, the action of the last term is trivial. 
We can express the action of $\nabla_A$ on such a tensor which is the representative of an AdS$_{d+1}$ tensor as:
\be
\nabla_B T_{A_1\dots A_r}(X) = G_B{}^C G_{A_1}{}^{C_1} \dots G_{A_r}{}^{C_r} \frac{\pa}{\pa X^C} T_{C_1\dots C_r}(X).
\ee
In particular, it is worth noting that induced AdS$_{d+1}$ metric $G_{AB}$ itself also satisfies transverse condition $X^A G_{AB} = G_{AB}X^B=0$, we have
\be
\nabla_C G_{AB} = G_C{}^{C'} G_{A}{}^{A'} G_{B}{}^{B'} \frac{\pa}{\pa X^{C'}}G_{A' B'} = 0
\ee 
as  required for $\nabla_A$ to be the metric covariant derivative in the embedding space.

\section{Integrals for Three Point Geodesic Witten Diagrams}\label{App:Integral}
\subsection*{Scalar Integral}
\paragraph{}
Here we compute the integral associated with the three point scalar geodesic Witten diagram, 
which is frequently used in the main text:
\begin{eqnarray}
\mathcal{A}^{\Delta_1\Delta_2 \Delta_0}_3(P_1,P_2,P_0)\equiv \int^\infty_{-\infty}\! d\lambda \frac{1}{(-2P_1\cdot {X}(\lambda))^{\Delta_1}} \frac{1}{(-2P_2\cdot {X}(\lambda))^{\Delta_2}}\frac{1}{(-2P_0\cdot {X}(\lambda))^{{\Delta_0}}}\,.
\end{eqnarray}
where $X(\lambda)$ is given in \eqref{Def:X-geodesics}.
This can be computed readily using the integral definition of Beta function ${\rm B}(a, b) = \frac{\Gamma(a)\Gamma(b)}{\Gamma(a+b)}$, after the direct substitution of geodesic coordinate, 
we can express the integral above as:
\begin{eqnarray}
\label{calc-A}
\mathcal{A}^{\Delta_1\Delta_2\Delta_0}_3(P_1,P_2,P_0)&=&P_{12}^{-\frac{1}{2}(\Delta_1+\Delta_2-{\Delta_0})}P_{10}^{-{\Delta_0}}
\int_{-\infty}^\infty\! d\lambda ~e^{(-\Delta_1+\Delta_2+{\Delta_0})\lambda}\left(\frac{P_{20}}{P_{10}}\,e^{2\lambda}+1\right)^{-{\Delta_0}}\no\\
&=&P_{12}^{-\frac{1}{2}(\Delta_1+\Delta_2-{\Delta_0})}P_{10}^{-{\Delta_0}}\left(\frac{P_{10}}{P_{20}}\right)^{\frac{1}{2}(-\Delta_1+\Delta_2+{\Delta_0})}
\int_0^\infty\! \frac{d\tilde{t}}{2\tilde{t}}~\tilde{t}^{\frac{1}{2}(-\Delta_1+\Delta_2+{\Delta_0})}\left(\tilde{t}+1\right)^{-{\Delta_0}}\no\\
&=& \frac{\beta_{\Del_{12}, \Del_0}}{P_{12}^{\frac{1}{2}(\Delta_1+\Delta_2-{\Delta_0})}P_{20}^{\frac{1}{2}(\Delta_2+{\Delta_0}-\Delta_1)}P_{10}^{\frac{1}{2}({\Delta_0}+\Delta_1-\Delta_2)}},
\end{eqnarray}
where
\begin{eqnarray}
\beta_{\Del_{12}, \Del_0}\equiv\frac{1}{2}{\rm B}\left(\frac{\Del_0+\Delta_{12}}{2},\frac{\Del_0-\Delta_{12}}{2}\right)
=\frac{\Gamma\left(\frac{\Del_0+\Delta_{12}}{2}\right)\Gamma\left(\frac{\Del_0-\Delta_{12}}{2}\right)}{2\Gamma({\Delta_0})}\,.
\end{eqnarray}
In the second line in (\ref{calc-A}), we have made the following change of integration variable: $\tilde{t} = \frac{P_{20}}{P_{10}} e^{2\lambda}$.

\subsection*{Spin-$l$ Integral}
\paragraph{}
Here we consider the spin-$l$, $l=0, 1, \dots J-1$ generalization of the computation for spin-$J$ case done in \eqref{3pt-ScalarGWD-J}.
The corresponding three point interaction vertex is:
\be\label{Scalar-3pt-Int-vertex-l}
g_{\Phi_1\Phi_2\Xi_l}\int_{X=X(\lambda)} dX \nabla^{C_1}\dots \nabla^{C_r}\Phi_1(X) \nabla^{C_{r+1}} \dots \nabla^{{C_J}} \Phi_2(X)\nabla_{C_1}\dots \nabla_{C_{J-l}} \Xi(X)_{C_{J-l+1}\dots C_J}.
\ee
The vertex \eqref{Scalar-3pt-Int-vertex-l} generates the following three point geodesic Witten diagram:
\bea\label{3pt-ScalarGWD-l}
&&\int_{\gamma_{12}} (K\cdot\nabla)^r\frac{\cC_{\Del_1}}{(-2P_1\cdot X)^{\Del_1}} (K\cdot\nabla)^{J-r}  \frac{\cC_{\Del_2}}{(-2P_2\cdot X)^{\Del_2}} 
(W\cdot \nabla)^{J-l}  \left[  \cC_{h+i\nu, l}  \frac{(2 X\cdot C_0\cdot W)^{l}}{(-2P_0\cdot X)^{h+i\nu+l}}\right]\nn\\
 && =\cP_{\Del_1, \Del_2, h+i\nu}^{l, J-r}  \int_{\gamma_{12}}  \frac{ (2P_1\cdot G \cdot K)^r (2P_2\cdot G \cdot K)^{J-r}}{(-2P_1\cdot X)^{\Del_1+r} (-2P_2\cdot X)^{\Del_2+J-r} }  
   \left[\frac{(2P_0\cdot W)^{J-l} (2 X\cdot C_0\cdot W)^{l}}{(-2P_0\cdot X)^{h+i\nu+l}}\right]\nn\\
  &&= \cP_{\Del_1, \Del_2, h+i\nu}^{l, J-r} (-1)^{J-r}  \int^{\infty}_{-\infty} d\lambda\frac{    \left(\frac{dX(\lam)}{d\lambda}\cdot K\right)^J}{  (-2P_1\cdot X(\lambda))^{\Del_1} (-2P_2\cdot X(\lam))^{\Del_2}}
  \frac{(2P_0\cdot W)^{J-l} (2X(\lam)\cdot C_0\cdot W)^{l}}{(-2P_0\cdot X(\lam))^{h+i\nu+J}}\nn\\
  &&=  \cP_{\Del_1, \Del_2, h+i\nu}^{l, J-r} (-1)^{J-r} J!\left(\frac{d-1}{2}\right)_J\int^{\infty}_{-\infty} d\lam \frac{ [2 {\rm V}_{0,12}]^{l} \left(\frac{(-2P_0\cdot P_2)}{(-2P_2\cdot X(\lambda))} -  \frac{(-2P_0\cdot P_1)}{(-2P_1\cdot X(\lambda))}   \right)^{J-l}}{(-2P_1\cdot X(\lambda))^{\Del_1} (-2P_2\cdot X(\lam))^{\Del_2}  (-2P_0\cdot X(\lam))^{h+i\nu+J} }\nn\\
&&= 2^{l}(-1)^{J-r} J!\left(\frac{d-1}{2}\right)_J{\mathbb P}_{\Del_1, \Del_2, h+i\nu}^{l, J-r}  \cC_{h+i\nu, l} \beta_{ \Del_{12}, h+i\nu+l } 
\begin{bmatrix}
				\Delta _1 &\Delta _2&  h+i\nu       \\[0.05em]
				0 & 0           &  l \\[0.05em]
				0         &     0     & 0  
			\end{bmatrix}		
\eea
The overall factor is defined to be:
\be\label{Def:CP-factor}
\cP_{\Del_1, \Del_2, \frac{d}{2}+i\nu}^{l, J-r}= \cC_{\Del_1}\cC_{\Del_2} \cC_{h+i\nu, l}(\Del_1)_r(\Del_2)_{J-r}    \left(h+i\nu+l\right)_{J-l},
\ee
\be\label{Def:bbP-factor}
{\mathbb P}_{\Del_1, \Del_2, h+i\nu}^{l, J-r}=   \cC_{\Del_1}\cC_{\Del_2} (\Del_1)_r(\Del_2)_{J-r}  \sum_{s=0}^{J-l} \frac{(J-l)! (-1)^{J-l-s}}{s! (J-l-s)!} 
{\left(\frac{h+i\nu+l-\Del_{12}}{2}\right)_s \left(\frac{h+i\nu+l+\Del_{12}}{2}\right)_{J-l-s}}.
\ee
Up to an overall factor $\bbP^{l, J-r}_{\Del_1, \Del_2, h+i\nu}$, which despite its dependence on $\nu$, does not introduce additional singularities for $\nu$ integration, 
we see that spin-$l$ case \eqref{3pt-ScalarGWD-l} takes exactly same expression for its spin-$J$ counterpart \eqref{3pt-ScalarGWD-J} with trivial substitution $J \to l$.

\section{Integrals for Three Point normal Witten Diagrams}\label{App:normal-Witten}
\paragraph{}
Here we consider the integration of normal three point Witten diagram with scalar fields.
The following calculation is based on \cite{PenedonesMellinAmp}:
\begin{eqnarray}\label{3pt-scalar-Witten}
I^{\text{3-pt}}\equiv\int\! dX \frac{1}{(-2P_1\cdot X)^{\Delta_1}}\frac{1}{(-2P_2\cdot X)^{\Delta_2}}\frac{1}{(-2P_3\cdot X)^{\Delta_3}}\,.
\end{eqnarray} 
Using the Schwinger parameterization, 
\begin{eqnarray}
\frac{1}{(-P_i\cdot X)^{\Delta_i}}=\frac{1}{\Gamma(\Delta_i)}\int^{\infty}_{0} \! \frac{dt_i}{t_i} ~t^{\Delta_i}
e^{-(-2P_i\cdot X)t_i}\,,
\end{eqnarray} 
we can rewrite the integration as
\begin{eqnarray}
I^{\text{3-pt}}=\frac{1}{\Gamma(\Delta_1)\Gamma(\Delta_2)\Gamma(\Delta_3)}\int^\infty_{0}\! 
\frac{dt_1}{t_1} \frac{dt_2}{t_2} \frac{dt_3}{t_3} 
t_1^{\Delta_1}t_2^{\Delta_2}t_3^{\Delta_3}\int\! dX e^{2Q\cdot X}\,,
\end{eqnarray} 
where $Q$ is defined as $Q\equiv\sum_{i=1}^3 t_i P_i$\,. 
Because $Q\cdot X$ is a scalar under the Lorentz transformation
in the embedding space $\mathbb{M}^{d+1,1}$\,, we can choose $Q$ as $|Q|(1,1,0)$ 
where $|Q|^2=\sum_{i>j} t_i t_j P_{ij}$\,.
Now the coordinate $X$ is parametrized as $(1,z^2+y^2,y^{\mu})/z$, we can evaluate the AdS integration
\begin{eqnarray}
\int\! dX e^{2Q\cdot X} &=&\int^\infty_0\!\frac{dz}{z} \int_{\mathbb{R}^d}\! \frac{d^dy}{z^d}
e^{-\frac{|Q|}{z}(z^2+y^2+1)}\no\\
&=&\pi^h \int^\infty_0\!\frac{dz}{z}  \frac{1}{(z |Q|)^h} e^{-\frac{|Q|}{z}(z^2+1)}\no\\
&=&\pi^h \int^\infty_0\!\frac{dz}{z}  \frac{1}{z^h} e^{-\left(z+\frac{|Q|^2}{z}\right)}\,,
\end{eqnarray} 
in the last line, $z$ is scaled as $z\rightarrow |Q|^{-1} z$\,.
Scaling $t_i$ as $t_i\rightarrow t_i \sqrt{z}$, we can perform the $z$ integration
\begin{eqnarray}
I^{\text{3-pt}}&=&\frac{\pi^h}{\Gamma(\Delta_1)\Gamma(\Delta_2)\Gamma(\Delta_3)}\int^\infty_0\! 
\frac{dt_1}{t_1} \frac{dt_2}{t_2} \frac{dt_3}{t_3} 
\int^\infty_0 \! \frac{dz}{z}z^{\frac{\Delta_1+\Delta_2+\Delta_3-d}{2}}e^{-z-|Q|^2}\no\\
&=&\frac{\pi^h}{\Gamma(\Delta_1)\Gamma(\Delta_2)\Gamma(\Delta_3)}
\Gamma\left(\frac{\sum_{i=1}^3\Delta_i-d}{2}\right)\int^\infty_0\! 
\frac{dt_1}{t_1} \frac{dt_2}{t_2} \frac{dt_3}{t_3} t_1^{\Delta_1}t_2^{\Delta_2}t_3^{\Delta_3} e^{-\sum_{i>j} t_i t_j P_{ij}}\,.
\end{eqnarray} 
By utilizing the following parameterization:
\begin{eqnarray}
t_1=\sqrt{\frac{m_1 m_3}{m_2}}\,,\qquad t_2=\sqrt{\frac{m_1 m_2}{m_3}}\,,\qquad  t_3=\sqrt{\frac{m_2 m_3}{m_1}}\,,
\end{eqnarray} 
the $t_i$ integration can be calculated as 
\begin{eqnarray}
\int^\infty_{0}\! 
&&\frac{dt_1}{t_1} \frac{dt_2}{t_2} \frac{dt_3}{t_3} t_1^{\Delta_1}t_2^{\Delta_2}t_3^{\Delta_3} e^{-\sum_{i>j} t_i t_j P_{ij}}\\
&&\quad=\frac{1}{2}\int^{\infty}_0\! \frac{dm_1}{m_1}\frac{dm_2}{m_2}\frac{dm_3}{m_3}
m_1^{\frac{\Delta_1+\Delta_2-\Delta_3}{2}}m_2^{\frac{\Delta_2+\Delta_3-\Delta_1}{2}}
m_3^{\frac{\Delta_3+\Delta_1-\Delta_2}{2}}e^{-m_1 P_{12}-m_2 P_{23}-m_3P_{31}}\no\\
&&\quad=\frac{1}{2}\Gamma\left(\frac{\Delta_1+\Delta_2-\Delta_3}{2}\right)
 \Gamma\left(\frac{\Delta_1-\Delta_2+\Delta_3}{2}\right)\Gamma\left(\frac{\Delta_2-\Delta_1+\Delta_3}{2}\right)\no\\
 &&\qquad \times  P_{12}^{-\frac{1}{2}(\Delta_1+\Delta_2-\Delta_3)}P_{23}^{-\frac{1}{2}(\Delta_2+\Delta_3-\Delta_1)}P_{31}^{-\frac{1}{2}(\Delta_3+\Delta_1-\Delta_2)}\,.
 \end{eqnarray} 
 Therefore the three point scalar diagram \eqref{3pt-scalar-Witten} can be evaluated as
 \begin{eqnarray}
 I^{\text{3-pt}}=\mathcal{N}^{\Delta_1,\Delta_2, \Delta_3}\hat{\cA}^{\Delta_1,\Delta_2,\Delta_3}\,,
  \end{eqnarray} 
  where
 \begin{eqnarray}
 &&\mathcal{N}^{\Delta_1,\Delta_2, \Delta_3}\equiv\\ 
&&\quad \frac{\pi^h\Gamma\left(\frac{\sum^3_{i=1}\Delta_i-d}{2}\right)}{2 \Gamma(\Delta_1)\Gamma(\Delta_2)\Gamma(\Delta_3)}
\Gamma\left(\frac{\Delta_1+\Delta_2-\Delta_3}{2}\right)
 \Gamma\left(\frac{\Delta_1-\Delta_2+\Delta_3}{2}\right)\Gamma\left(\frac{\Delta_2-\Delta_1+\Delta_3}{2}\right)\,,\label{Def:N123}\no\\
&&\hat{\cA}^{\Delta_1,\Delta_2,\Delta_3}\equiv
 P_{12}^{-\frac{1}{2}(\Delta_1+\Delta_2-\Delta_3)}P_{23}^{-\frac{1}{2}(\Delta_2+\Delta_3-\Delta_1)}P_{31}^{-\frac{1}{2}(\Delta_3+\Delta_1-\Delta_2)}\,.
\end{eqnarray}

\section{Rewriting tensor structures and some useful identities}\label{rewritting-ts}
\paragraph{}
In this appendix we consider more explicit proof of the statement that the three point geodesic Witten diagrams involving spins, formed by arbitrary Lorentz invariant vertices, 
can be expressed in terms of linear combination of box tensor basis, filling in some details for the general arguments given in \cite{SpinningBlock1}.
\paragraph{}
Here we show that a transverse polynomial $\bbQ(P_i,Z_i)$ ($i=1,2,3$) be built only from $\rH_{ij}$
and $\rV_{i,jk}$\,.
We assume that the polynomial $\bbQ(P_i,Z_i)$ has degree $l_i$ in $Z_i$ and it is transverse in each $Z_i$, in other wards, $\bbQ(P_i,Z_i)$ is invariant under the following shift of $Z_i$;
\begin{eqnarray}
Z_i\rightarrow Z_i+\alpha_i P_i\,,
\end{eqnarray}
where $\alpha_i$ are arbitrary constants. 
Because this polynomial $\bbQ$ do not have the Lorentz indices, $\bbQ$ can only consist of three scalar products; $P_i\cdot P_j$, $Z_i\cdot P_j$ and $Z_i\cdot Z_j$\,.
The combination $Z_i\cdot Z_j$ to $\rH_{ij}$ is replaced to $\rH_{ij}$ and other scalar product through (\ref{Def:Hij}).
Then $\bbQ$ can be represented as;
\begin{eqnarray}
\bbQ(P_i, Z_i)=\sum_{(m_1,m_2,m_3)=(0,0,0)}^{(l_1,l_2,l_3)}R_{m_1,m_2,m_3}((P_i\cdot P_j) \,,(Z_i\cdot P_j) ,\, \rH_{ij})\,,
\end{eqnarray}
where $R_{m_1,m_2,m_3}$ is a polynomial consisted of $P_i\cdot P_j$, $Z_i\cdot P_j$ and $\rH_{ij}$ and it contains $m_i$ $Z_i$ besides $\rH_{ij}$.
We can decompose $R_{m_1,m_2,m_3}$ further;
\begin{eqnarray}
R_{m_1,m_2,m_3}=\sum_{n=0}^{m_1}c_{n,m_1-n}\,(Z_1\cdot P_2)^{n}\,(Z_1\cdot P_3)^{m_1-n}\,.
\label{exp-R}
\end{eqnarray}
Here we focus on the specific $Z_1$ dependence. 
The coefficient $c_{n,m_1-n}$ depends on $Z_1$ only through $\rH_{12}$ or $\rH_{31}$\,.
$\bbQ$ should satisfy the transverse condition;
\begin{eqnarray}
\bbQ(P_i, Z_i)=\bbQ(P_i, Z_i+\beta_i P_i)\,,
\end{eqnarray}
therefore the following equation should be satisfied 
\begin{eqnarray}
\frac{\partial}{\partial \beta}\left[\sum_{n=0}^{m_1}c_{n,m_1-n}(Z_1\cdot P_2+\beta P_1\cdot P_2)^{n}(Z_1\cdot P_3+\beta P_1\cdot P_3)^{m_1-n}\right]=0\,.
\end{eqnarray}
Because this condition should satisfied at each order of $(Z_1\cdot P_2)$ or $(Z_1\cdot P_2)$, 
we can obtain the following recursion equation;
\begin{eqnarray}
(m_1-n)\,c_{n,m_1-n}(P_1\cdot P_3)+(n+1)\,c_{n+1,m_1-n-1}(P_1\cdot P_2)=0\,.
\end{eqnarray}
According to this relation, $c_{n,m_1-n}$ is determined as
\begin{eqnarray}
c_{n,m_1-n}={}_{m_1}C_n\, \left(-\frac{P_1\cdot P_3}{P_1\cdot P_2}\right)^n c_{\,0,m_1}\,.
\end{eqnarray}
Then the decomposition in (\ref{exp-R}) is just a binomial expansion and $R_{m_1,m_2,m_3}$ can be rewritten as;
\begin{eqnarray}
R_{m_1,m_2,m_3}&=&c_{\,0,m_1}  \left(-\frac{P_2\cdot P_3}{P_1\cdot P_2}\,\rV_{1,23}\right)^{m_1}\,.
\end{eqnarray}
The discussions for $Z_2$ and $Z_3$ go through similarly.
Therefore transverse polynomials $\bbQ$ should depend on depends on $Z_i$ only through
$\rH_{ij}$ and $\rV_{i,jk}$\,.
\paragraph{}
 Here we also list out few useful identities which involve the contractions among triplets of anti-symmetric $\rC_i^{AB}$ associated with $\cO_{\Del_i, l_i}(P_i, Z_i)$, $i=1,2,3$, 
 which are useful in the actual explicit computations:
\bea
\label{Def:2c}
\left(\rC_i \cdot \rC_j\right)_{AB}&=&
\frac{1}{2}\frac{{\rH}_{ij}}{\left(P_i \cdot  P_j\right)}P_{iA}P_{jB}
-\frac{\left(P_j\cdot \rC_i\right)_A\left(P_i \cdot \rC_j\right)_B}{\left(P_i  \cdot P_j\right)},\\
\label{Def:3c}			
\left(\rC_i  \cdot \rC_j  \cdot \rC_k\right)_{{AC}}&=&
\frac{{\rH}_{ij}}{2}\frac{P_{iA}\left(P_j  \cdot \rC_k\right)_C}{\left(P_i  \cdot P_j\right)}
-\frac{{\rH}_{jk}}{2}\frac{P_{kC}\left(P_j \cdot \rC_i\right)_A}{\left(P_i  \cdot P_j\right)}\no\\
&&+\frac{\left(P_j  \cdot \rC_i\right)_A\left(P_j  \cdot \rC_k\right)_C}{\left(P_i  \cdot P_j\right)\left(P_j  \cdot P_k\right)}\left(P_k \cdot \rC_j  \cdot P_i\right),\\ 
\left(\rC_i \cdot \rC_j \cdot \rC_i\right)_{AC}&=&-\frac{{\rH}_{ij}}{2}\rC_{iAC}\\
\label{Def:4c1}
\left(\rC_i \cdot  \rC_j \cdot  \rC_k \cdot  \rC_j\right)_{AD}&=&
\frac{-{\rH}_{ij}{\rH}_{jk}}{4}\frac{P_{iA}P_{jD}}{\left(P_i \cdot  P_j\right)}
+\frac{{\rH}_{{jk}}\left(P_j \cdot   \rC_i\right)_A\left(P_i \cdot   \rC_j\right)_D}{2\left(P_i \cdot   P_j\right)},\\	
\label{Def:4c2}						
\left(\rC_i \cdot  \rC_j \cdot  \rC_k \cdot  \rC_i\right)_{{AD}}&=&
\frac{{\rH}_{ij}{\rH}_{ki}}{4}\frac{\left(P_j\cdot P_k\right)P_{iA}P_{iD}}{\left(P_i  \cdot   P_j\right)\left(P_k  \cdot   P_i\right)}
-\frac{1}{4}\rV_{j, ki}\rV_{k, ij}\left(P_j  \cdot   \rC_i\right)_A\left(P_j  \cdot   \rC_i\right)_D\no\\
&&-\frac{1}{2}{\rH}_{jk}\frac{\left(P_j  \cdot  \rC_i\right)_A\left(P_k. \rC_j\right)_D}{\left(P_i  \cdot  P_j\right)} 							
-\frac{1}{2}{\rH}_{ij}\frac{\rV_{k, ij}P_{\text{iA}}\left(P_k \cdot    \rC_i\right)_D}{\left(P_k\cdot P_i\right)}\no\\
&&+\frac{1}{2}{\rH}_{ki}\frac{\rV_{j,ki}\left(P_j  \cdot  P_k\right)}{\left(P_i  \cdot  P_j\right)}\left(P_j \cdot   \rC_i\right)_AP_{iD}.
\eea
Along with the obvious identity $(\rC_i\cdot \rC_i)_{AB} = 0$, all other  successive contractions of $\rC_{i}^{AB}$ can obtained by repeatedly using these identities.
It should be clear from above that any invariant scalars constructed from contacting \eqref{Def:2c}-\eqref{Def:4c2} with either a pair of $P_i$ or a $\rC_{i AB}$ 
are all transverse polynomials and can all be expressed in terms products of $\rH_{ij}$ and $\rV_{i, jk}$ with coefficients only depending on $(P_i\cdot P_j)$.
These will be needed when we study the tensor structures of the three point geodesic Witten diagrams.

\section{Computational Details for Decomposition Analysis}\label{App:Decomposition}
\paragraph{}
In Section \ref{Section:Decomposition}, we have demonstrated that ordinary four point scalar Witten diagram can be written as a summation of four point scalar CPWs.
In this case, each CPW is proportional to a GWD, this leads us to the claimed results, here we present the computational details to see such decomposition. 
In \eqref{3-pt-WD-I-2}, we can replace $\hat{\mathcal{A}}$ with a geodesic diagram as follows;
\begin{eqnarray}\label{3ptWD-GWD}
I_{(J,l)}^{\Delta_1,\Delta_2,h+i\nu}
&=& \sum^{J-l}_{p=0}{}_{J-l}C_p\, \frac{(-2)^l(\Delta_2)_J (h+i\nu+l)_{J-l}}{(h+i\nu+J-l-p)_l} (-2 P_{20})^{J-l-p} 
 \mathcal{N}^{\Delta_1, \Delta_2 +J-p,h+i\nu+J-l-p}\no\\
&& \qquad \times (\rD_{02})^l \hat{\cA}^{\Delta_1, \Delta_2 +J-p,h+i\nu+J-l-p}\no\\
&=& \sum^{J-l}_{p=0}{}_{J-l}C_p\, \frac{(\Delta_2)_J (h+i\nu+l)_{J-l}}{(h+i\nu+J-l-p)_l} (-2)^{J-p} 
 \mathcal{N}^{\Delta_1, \Delta_2 +J-p,h+i\nu+J-l-p}\no\\
 && \qquad \times \left(\frac{\Delta_{12}+h+i\nu -l}{2}\right)_l (-V_{0,12})^l \mathcal{A}^{\Delta_1,\Delta_2,h+i\nu +l}\no\\
 &=& \sum^{J-l}_{p=0}\tilde{\mathcal{R}}_{\Delta_1,\Delta_2,h+i\nu}^{J,l,p} \Gamma\left(\frac{\Delta_1+\Delta_2+l-h\pm i\nu}{2}\right) \\
 &&\qquad\times\frac{1}{l!\, \left(\frac{d-1}{2}\right)_l}\int_{\gamma_{12}}\! d\lambda \frac{1}{(-2P_1\cdot X)^{\Delta_1}}(K\cdot \nabla)^l\frac{1}{(-2P_2\cdot X)^{\Delta_2}}
 \frac{(W\cdot V_0(X))^l}{(-2P_0\cdot X)^{h+i\nu +l}}\,,\no
 \end{eqnarray}
where in the last line we used the following relation:
\begin{eqnarray}
&& \frac{1}{l!\, \left(\frac{d-1}{2}\right)_l}\int_{\gamma_{12}}\! d\lambda \frac{1}{(-2P_1\cdot X)^{\Delta_1}}(K\cdot \nabla)^l\frac{1}{(-2P_2\cdot X)^{\Delta_2}}
 \frac{(W\cdot V_0(X))^l}{(-2P_0\cdot X)^{h+i\nu +l}}\no\\ 
&&\qquad  \qquad = (\Delta_2)_l (-2)^l \beta^{\Delta_1,\Delta_2, h+i\nu+l} (-V_{0,12})^l 
\mathcal{A}^{\Delta_1,\Delta_2,h+i\nu+l}\,,
 \end{eqnarray}
 and $\tilde{\mathcal{R}}_{\Delta_1,\Delta_2,h+i\nu}^{J,l,p}$ is a regular function of $\nu$:
 \begin{eqnarray}
 \tilde{\mathcal{R}}_{\Delta_1,\Delta_2,h+i\nu}^{J,l,p}&\equiv& {}_{J-l}C_p \frac{\pi^h (-2)^{J-l-p}(\Delta_2+l)_{J-l}}{\Gamma(\Delta_1)\Gamma(\Delta_2+J-p)} (h+i\nu +J-p)_p\\
 &&\qquad \times\left(\frac{-\Delta_{12}+h+i\nu+l}{2}\right)_{J-l-p}\left(\frac{\Delta_1+\Delta_2-h+i\nu+l}{2}\right)_{J-l-p} \,.\no
 \end{eqnarray}
Using \eqref{3ptWD-GWD}, now we can rewrite \eqref{4-pt-NWD-2} as:
\begin{eqnarray}\label{4pt-nuGWD}
W^{\text{4-pt.}}_{(0,0),J,(0,0)}&=&\sum^J_{l=0}\int_\partial \!dP_0 \int^\infty_{-\infty}\! d\nu~ a_l(\nu)\frac{\nu^2}{\pi l!\, (h-1)!}
\sum^{J-l}_{p=0}\sum^{J-l}_{p'=0}
\tilde{\mathcal{R}}_{\Delta_1,\Delta_2,h+i\nu}^{J,l,p}\tilde{\mathcal{R}}_{\Delta_3,\Delta_4,h-i\nu}^{J,l,p'}\\
&&\times\Gamma\left(\frac{\Delta_1+\Delta_2+l-h\pm i\nu}{2}\right) \Gamma\left(\frac{\Delta_3+\Delta_4+l-h\pm i\nu}{2}\right) \no\\
&&\times\frac{\mathcal{C}_{h+i\nu,l}}{l!\, \left(\frac{d-1}{2}\right)_l}\int_{\gamma_{12}}\! d\lambda \frac{1}{(-2P_1\cdot X)^{\Delta_1}}(K\cdot \nabla)^l\frac{1}{(-2P_2\cdot X)^{\Delta_2}}
 \frac{(W\cdot V_0(X,D_{Z_0}))^l}{(-2P_0\cdot X)^{h+i\nu +l}}\no\\
&&\times\frac{\mathcal{C}_{h-i\nu,l}}{l!\, \left(\frac{d-1}{2}\right)_l}\int_{\gamma_{34}}\! d\lambda' \frac{1}{(-2P_3\cdot \tilde{X})^{\Delta_3}}(\tilde{K}\cdot \tilde{\nabla})^l\frac{1}{(-2P_4\cdot \tilde{X})^{\Delta_4}}
 \frac{(\tilde{W}\cdot V_0(\tilde{X},Z_0))^l}{(-2P_0\cdot \tilde{X})^{h-i\nu +l}}\no\,.
\end{eqnarray}
From \eqref{Def:AdSHarm0} and \eqref{Def:AdSHarm}, 
the two bulk to boundary propagator can be glued together:
\begin{eqnarray}
&&\int_\partial \!dP_0 ~ a_l(\nu)\frac{\nu^2}{\pi l!\, (h-1)!} \mathcal{C}_{h+i\nu,l} \frac{(W\cdot V_0(X,D_{Z_0}))^l}{(-2P_0\cdot X)^{h+i\nu +l}} \mathcal{C}_{h-i\nu,l}\frac{(\tilde{W}\cdot V_0(\tilde{X},Z_0))^l}{(-2P_0\cdot \tilde{X})^{h-i\nu +l}}\no\\
&&\qquad =\frac{i \nu}{2\pi }\left(\Pi_{h+i\nu,l}(X,\tilde{X};W,\tilde{W})-\Pi_{h-i\nu,l}(X,\tilde{X};W,\tilde{W})\right)\,,
\end{eqnarray}
then \eqref{4pt-nuGWD} becomes
\begin{eqnarray}\label{4pt-nuGWD2}
W^{\text{4-pt.}}_{(0,0),J,(0,0)}&=&\sum^J_{l=0} \int^\infty_{-\infty}\! d\nu~ a_l(\nu) \frac{i \nu}{2\pi}
\sum^{J-l}_{p=0}\sum^{J-l}_{p'=0}
\tilde{\mathcal{R}}_{\Delta_1,\Delta_2,h+i\nu}^{J,l,p}\tilde{\mathcal{R}}_{\Delta_3,\Delta_4,h-i\nu}^{J,l,p'}\nn\\
&&\times\Gamma\left(\frac{\Delta_1+\Delta_2+l-h\pm i\nu}{2}\right) \Gamma\left(\frac{\Delta_3+\Delta_4+l-h\pm i\nu}{2}\right) \no\\
&&\times\left(\mathcal{W}_{h+i\nu,l}(P_i)-\mathcal{W}_{h-i\nu,l}(P_i)\right)\,,
\end{eqnarray}
where $\mathcal{W}_{\Delta,l}(P_i)$ is a four point GWD;
\begin{eqnarray}
\mathcal{W}_{\Delta,l}(P_i)&=& \frac{1}{\left(l!\, \left(\frac{d-1}{2}\right)_l\right)^2}\int_{\gamma_{12}}\! d\lambda \int_{\gamma_{34}}\! d\lambda'
\frac{1}{(-2P_1\cdot X)^{\Delta_1}}(K\cdot \nabla)^l\frac{1}{(-2P_2\cdot X)^{\Delta_2}}\no\\
&& \times\frac{1}{(-2P_3\cdot \tilde{X})^{\Delta_3}}(\tilde{K}\cdot \tilde{\nabla})^l\frac{1}{(-2P_4\cdot \tilde{X})^{\Delta_4}}
\left(\Pi_{h+i\nu,l}(X,\tilde{X};W,\tilde{W})-\Pi_{h-i\nu,l}(X,\tilde{X};W,\tilde{W})\right)\,.\nn\\
\end{eqnarray}
In \eqref{4pt-nuGWD2}, the intermediate states are determined by the pole structure of $\nu$ integration.
The $\nu$ dependence is similar as in \eqref{GWDandCB}, the single trace contribution comes form the highest spin coefficient $a_J(\nu)$,
and the double trace ones come from the gamma functions.
In this more direct computation, we can see the decomposition into GWDs without passing through CPWs.

\bibliographystyle{sort}

\end{document}